\documentclass[12pt,a4paper]{article}
\usepackage{jcappub}
\usepackage{newtxtext,newtxmath}
\usepackage{color,xcolor}
\usepackage{graphicx}	
\usepackage{latexsym,amsmath,amssymb}
\usepackage{multicol}
\usepackage{subfigure,float}
\usepackage{dblfloatfix}
\usepackage{soul}

\DeclareRobustCommand{\VAN}[3]{#2}
\let\VANthebibliography\thebibliography
\def\thebibliography{\DeclareRobustCommand{\VAN}[3]{##3}\VANthebibliography}

\usepackage{array}
\newcolumntype{P}[1]{>{\centering\arraybackslash}p{#1}}

\DeclareUnicodeCharacter{2009}{~}
\DeclareUnicodeCharacter{2248}{~}

\def \xhii {\bar{x}_{\rm HII}}

\def \HI{{\sc Hi}}

\def\tb{{\delta \bar{T}_{\rm b}}}

\def\fx{f_{\rm X}}
\def\mh{M_{\rm h,\, min}}
\def \mp{{\,\rm Mpc}^{-1}}

\newcommand{\TB}{\delta T_{\rm b}}

\newcommand{\OmegaB}{\Omega_{\rm B}}
\newcommand{\Omegam}{\Omega_{\rm m}}

\newcommand{\be}{\begin{equation}}
\newcommand{\e}{\end{equation}}
\newcommand{\bear}{\begin{eqnarray}}
\newcommand{\ear}{\end{eqnarray}}

\newcommand{\comment}[1]{}

\title{Redshifted 21-cm bispectrum: Impact of the source models on the signal and the 
IGM physics from the Cosmic Dawn}
\author[a,b]{Mohd Kamran,}\emailAdd{kamranmohd080@gmail.com}
\author[c,d]{Raghunath Ghara,}
\author[a,e]{Suman Majumdar,}
\author[f]{Garrelt Mellema,}
\author[g]{Somnath Bharadwaj,}
\author[e]{Jonathan R. Pritchard,}
\author[h]{Rajesh Mondal}
\author[i]{and Ilian T. Iliev}

\affiliation[a]{Department of Astronomy, Astrophysics \& Space Engineering, Indian Institute of Technology Indore, Indore 453552, India}
\affiliation[b]{Department of Physics and Astronomy, Uppsala University, Uppsala 75237, Sweden}
\affiliation[c]{ARCO (Astrophysics Research Center), Department of Natural Sciences, The Open University of Israel, 1 University Road, PO Box 808, Ra'anana 4353701, Israel}
\affiliation[d]{Department of Physics, Technion, Haifa 32000, Israel}
\affiliation[e]{Department of Physics, Blackett Laboratory, Imperial College, London SW7 2AZ, U. K.}
\affiliation[f]{The Oskar Klein Centre and The Department of Astronomy, Stockholm University, AlbaNova, SE-10691 Stockholm, Sweden}
\affiliation[g]{Department of Physics, Indian Institute of Technology Kharagpur, Kharagpur - 721302, India}
\affiliation[h]{School of Physics and Astronomy, Tel Aviv University, Tel Aviv 69978, Israel}
\affiliation[i]{Department of Physics and Astronomy, Pevensey II Building, University of Sussex, Brighton BN1 9QH, UK}

\date{\today}

\abstract{
The emissions from the first luminous sources drive the fluctuations in the 21-cm signal at Cosmic Dawn (CD) via two main astrophysical processes, namely Ly$\alpha$ coupling and X-ray heating, yielding a highly non-Gaussian signal. The impact of these processes on the 21-cm signal and its non-Gaussianity depends on the properties of these first sources of light. In this work, we consider different CD scenarios by varying two major source parameters i.e. the minimum halo mass $\mh$ and X-ray photon production efficiency $\fx$ using the 1D radiative transfer code {\sc grizzly}. We study the impact of variation in these source parameters on the large scale ($k_1 = 0.16 \mp$) 21-cm bispectrum for all possible unique triangles in the Fourier domain. Our detailed and comparative analysis of the power spectrum and bispectrum shows that the shape, sign and magnitude of the bispectrum jointly provide a better measure of the signal fluctuations and its non-Gaussianity than the power spectrum alone. We also conclude that it is important to study the sequence of sign changes along with the variations in the shape and magnitude of the bispectrum throughout the CD history to arrive at a robust conclusion about the dominant processes in the intergalactic medium at different cosmic times. We further observe that among all the possible unique $k$-triangles, the large-scale non-Gaussianity of the signal is best probed by the small $k$-triangles in the squeezed limit and by triangle shapes in its vicinity. This opens up the possibility of constraining the source parameters during the CD using the 21-cm bispectrum.
}


\begin{document}
\maketitle

\section{Introduction}
\label{intro}
The Cosmic Dawn and Epoch of Reionization (CD-EoR) is the period in the Universe's cosmic history during which the first light from luminous sources such as stars, galaxies, quasars, etc. appeared. The emissions from these first sources affected profoundly the state of the Universe by completely transforming the thermal and ionization states of the intergalactic medium (IGM) during this period. A considerable theoretical understanding of this epoch has been developed, but so far this has been little tested through observations. The redshifted 21-cm signal, produced by the spin-flip transition of the electron-proton system in the 1s ground state of the neutral hydrogen (\HI), is a direct tracer of \HI\ distribution in the IGM, observed through low-frequency radio observations. This signal also carries a wealth of information about the first sources, and their varying properties, which drive the fluctuations in the 21-cm signal. This opens up the possibility to use such observations to learn about these sources and the range of underlying astrophysical processes \citep{barkana01, furlanetto06, pritchard12}. The likely astrophysical processes dominant during the CD-EoR are the Ly$\alpha$ coupling, X-ray heating, and photo-ionization of the IGM gas. It is widely accepted that the stars in the galaxies are the major sources of UV photons that ionize most of the \HI\ in the IGM and provide the Ly-$\alpha$ background. In contrast, quasars and X-ray binaries (XRBs), are believed to be the primary sources of the X-ray photons that drive the IGM heating. The dominant astrophysical processes decide the nature and amplitude of the 21-cm signal fluctuations around the sources and their evolution with time during the CD-EoR.

In order to detect the CD-EoR 21-cm signal, a number of observational efforts are ongoing with the first generation of radio interferometers such as the GMRT\footnote{\url{http://www.gmrt.ncra.tifr.res.in}} \citep{paciga13}, LOFAR\footnote{\url{http://www.lofar.org/}} \citep{mertens20}, MWA\footnote{\url{http://www.mwatelescope.org/}} \citep{barry19}, PAPER \citep{kolopanis19} and HERA\footnote{\url{https://reionization.org/}} \citep{deboer17,HERA21}. However, a number of observational obstacles such as foregrounds \cite{2016ApJ...818..139T, 2021MNRAS.500.2264H}, systematics \cite{2022MNRAS.509.3693M, 2022arXiv220302345G}, etc introduce difficulties in the detection of this signal, thereby these instruments are unable to image the CD-EoR maps and hope to have potential to detect the expected signal statistically, using different statistical measures such as variance \citep{iliev08, patil14, watkinson15}, power spectrum \citep{ali08, ghosh12, mertens20} etc.. So far, they have only been able to impose upper limits of the signal fluctuations using the spherically-averaged power spectrum \citep{barry19, li19, kolopanis19, mertens20, trott20, HERA21}. The upcoming Square Kilometre Array (SKA)\footnote{\url{http://www.skatelescope.org/}} \citep{mellema15, koopmans15, ghara16} will be able to go much further and provide also detailed tomographic images of the CD-EoR.

The spherically averaged power spectrum as an estimator fully characterizes a signal statistically only if it is a Gaussian random field \citep{bharadwaj04,bharadwaj05, barkana05, datta07a,datta14, mesinger07, lidz08, choudhury09b, mao12, jensen13, majumdar13, majumdar14, mondal15, mondal16, majumdar16, mondal17}. However, the cumulative effect of radiation source clustering, non-uniform heating, and ionization of the IGM gas by those sources introduce a high level of non-Gaussianity in the CD-EoR 21-cm maps \citep{bharadwaj05a, mellema06, mondal15, kamran20, mondal21, kamran21}. This non-Gaussianity evolves together with the progress in ongoing astrophysical processes \citep{mondal16, mondal17, shaw19, shaw20, majumdar18, majumdar20, kamran20}. By definition, the power spectrum provides the auto-correlation between the signal at a single Fourier mode and thus cannot capture this non-Gaussianity. The fundamental statistics which can capture this non-Gaussianity are skewness and kurtosis \citep{harker09, watkinson14, watkinson15, shimabukuro15a, kubota16, 2019MNRAS.487.1101R, ross21}. However, these statistics can only probe the non-Gaussian features at a single length scale. One requires some robust higher-order statistics such as the bispectrum to study how the non-Gaussian features at different length scales evolve \citep{shimabukuro15, majumdar18, majumdar20, kamran20, mondal21, kamran21}. 

The bispectrum, by definition, correlates the signal fluctuations in the Fourier space at three wave numbers ($k$ modes) when forming a closed triangle. This statistic can thus capture the non-Gaussianity present in the 21-cm maps. Furthermore, the bispectrum sign provides additional information, compared to the power spectrum, which is always positive \cite{bharadwaj05a,watkinson17,majumdar18,majumdar20}. These features make the bispectrum a robust statistical probe of all those additional fundamental characteristics of the signal that are genuinely immersed in the signal's non-Gaussianity. For instance, the IGM heating and Ly-$\alpha$ fluctuations, which are the primary source of the non-Gaussianity during the CD-EoR, are directly connected with the properties of the sources of radiation formed during this period. Therefore, one expects the bispectrum to be a potential probe of this IGM physics \cite{kamran21}. Furthermore, the nature and level of the non-Gaussianity in the 21-cm signal are expected to depend on the types of sources of radiation and the rate at which a particular source emits radiation, which in turn is intimately linked with the source parameters. Thus, it is expected that the bispectrum can potentially identify and distinguish the various kinds of CD-EoR sources \cite{watkinson19,kamran20,ma21}. Finally, if the bispectrum is sufficiently sensitive to the source parameters, it may put much tighter constraints on the CD-EoR parameters than the power spectrum alone \cite{shimabukuro16b,tiwari22,watkinson22}.

In earlier work \cite{kamran20}, we considered various sources of light during the CD, such as the star-forming galaxies, mini quasars (mini-QSOs), and high mass X-ray binaries (HMXBs). We showed that the evolution of the 21-cm bispectrum magnitude and the sign can distinguish these sources better than the 21-cm power spectrum does. In \citep{kamran21}, we showed how the 21-cm signal bispectrum probes the impact of all possible astrophysical processes on the signal fluctuations by capturing the intrinsic non-Gaussianity in the signal during CD. However, in \citep{kamran21}, we did not investigate how the range of possible X-ray sources, may impact the 21-cm signal and its bispectrum through the possible astrophysical processes. We have also not studied the effect of different $\mh$ (minimum mass of source hosting halos) values on the 21-cm signal and its bispectrum. These CD scenarios impact the IGM through the varying Ly$\alpha$ coupling, X-ray heating, and photo-ionization processes. In the current work we study how these source parameters impact the CD signal bispectrum. We consider several simulated CD scenarios corresponding to a range of possible combinations of the values of the related source parameters that directly control the physical processes going on in the IGM. We focus on how these different CD scenarios that affect the 21-cm signal via the different astrophysical processes will impact the large-scale 21-cm bispectrum. Additionally, we are interested in investigating what additional information the bispectrum for a variety of triangles in the Fourier space can draw compared to the bispectrum for a particular type of $k$-triangle.

The structure of the paper is as follows: In Section \ref{sec:sim} we present the simulations used to generate the 21-cm maps. Section \ref{sec:bis} describes our formalism for estimating the bispectrum from the simulated signal. Section \ref{sec:results} discusses our analysis of the bispectra for different source models. Finally in Section \ref{sec:summary} we summarise our results.

In this paper, we have used the cosmological parameters $h = 0.7$, $\Omega_{\mathrm{m}} = 0.27$, $\Omega_{\Lambda} = 0.73$, $\Omega_{\mathrm{b}} = 0.044$ consistent with the \textit{WMAP} results \citep{hinshaw13} and within the error bars consistent with the \textit{Planck} results \citep{planck14}.

\section{Simulation of the Cosmic Dawn 21-cm signal}
\label{sec:sim}
The redshifted 21-cm signal from the CD is measured against the Cosmic Microwave Background Radiation (CMBR). The strength of this radio signal is often described by the differential brightness temperature ($\TB$), which can be expressed as \citep{pritchard08}
\begin{equation}
\begin{split}
    \delta T_{\rm b}(\textbf{r},z) = 27x_{\rm HI}(\textbf{r},z)\big(1+\delta_{\rm b}(\textbf{r}, z)\big)\Big(1-\frac{T_{\rm CMB}(z)}{T_{\rm S}(\textbf{r},z)}\Big) \Big( \frac{\Omega_{\rm b}h^2}{0.023}\Big) \Big(\frac{0.15}{\Omega_{\rm m}h^2} \frac{1+z}{10}\Big)^{1/2}  {\, \rm mK} 
    \label{eq:tb}
\end{split}
\end{equation} 
where $\textbf{r}$ and $z$ denote the position and redshift of the 21-cm signal emitting region. The quantities $\delta_{\rm b}$ and $\, T_{\rm CMB} (z)$ denote the baryonic density contrast, and the CMBR temperature, respectively. In addition to the dependence on the density field and cosmological parameters (as shown in Equ.\ref{eq:tb}), $ \delta T_{\rm b}(\textbf{r},z)$ also depends on the redshift-space distortion effects due to the peculiar velocities of the gas in the IGM. The astrophysical imprints on $ \delta T_{\rm b}(\textbf{r},z)$ comes from the neutral fraction $x_{\rm HI}(\textbf{r},z)$ and the spin temperature $T_{\rm S}$ terms.  The \HI~ spin temperature $T_{\rm S}$ during CD can be expressed as  \citep{field58}
\begin{equation}
    {T_{\rm S}(\textbf{r},z)} = \frac{T_{\rm CMB}(z)+x_{\alpha}(\textbf{r},z)T_{\rm g}(\textbf{r},z)}{1+x_{\alpha}(\textbf{r},z)}
    \label{eq:ts}
\end{equation}
where $x_{\alpha}(\textbf{r},z)$ is the Ly$\alpha$ coupling coefficient and $T_{\rm g}(\textbf{r},z)$ is the IGM gas temperature. In Equ.\ref{eq:ts} we ignored the effect of collisional coupling of the 21-cm line due to encounters with other hydrogen atoms or electrons, since the contribution of that process is largely negligible in the IGM at the redshifts of interest here, although it does contribute inside haloes as well as at higher reshifts \citep{2002ApJ...572L.123I}. Note that we also do not consider any excess radio background \citep[e.g.,][]{2020MNRAS.498.4178M, 2021MNRAS.503.4551G, 2022JCAP...03..055G} to the CMBR.

\begin{table}
\begin{center}
\begin{tabular}{|l||*{4}{c|}}\hline
$f_{\rm X}$ &$0.1$&$2.15$&$46.4$&$1000$\\\hline
$M_{\rm h, Min}\, (10^9\, M_{\odot})$ &$1.0$&$2.15$&$4.64$&$10.0$\\\hline
\end{tabular}
\caption{Different values of the X-ray heating efficiency parameter $f_{\rm X}$ and the minimum mass of halo with radiating sources  $\mh$ as considered in this study. We select all possible pairs of the two source parameter ($\fx$, $\mh$) values and explore sixteen simulated CD scenarios in this study.}
\label{table1}
\end{center}
\end{table}

We use the {\sc grizzly} \citep{ghara15} code to simulate $ \delta T_{\rm b}$ maps at 23 redshift snapshots for different astrophysical source models in the redshift range between 10 and 18. The algorithm uses a one-dimensional radiative transfer scheme which approximates the photon transfer from individual sources as isotropic. We refer the reader to \citep{ghara15,ghara18} for details on the algorithm. {\sc grizzly} requires as input the dark matter halo catalogs, and the gas density and velocity fields interpolated on an uniform grid. The cosmological structures used in this study are taken from results of the PRACE\footnote{Partnership for Advanced Computing in Europe: \url{http://www.prace-ri.eu/}} project {\sc PRACE4LOFAR}. The dark-matter-only N-body simulation was performed in a volume $(500h^{-1})^3$ comoving Mpc$^3$ \citep[for details, see e.g.,][]{2019JCAP...02..058G} using {\sc cubep$^3$m} code \citep{Harnois12}. The outputs density and velocity fields are interpolated  on $600^3$ grids. The dark-matter halos are identified by an on-the-fly spherical overdensity halo finder, which here reliably resolves halos with masses higher than $\approx 10^9 $ M$_{\odot}$. 

The source model used in the {\sc grizzly} algorithm assumes the stellar mass ($M_{\ast}$) of a galaxy formed in a dark matter halo of mass $M_{\rm h}$ is $M_{\ast} = f_{\ast} \Big( \frac{\OmegaB}{\Omegam} \Big) M_{\rm h}$ where we fix the value of the star formation efficiency $f_{\ast}$ to 0.03 \citep{behroozi15,sun16} in this study. The algorithm also requires the spectral energy distribution (SED) per stellar mass of a galaxy as input for the 1D radiative transfer. The SED per stellar mass of a galaxy used in the code is generated using a publicly available population synthesis code {\sc pegase2} \citep{Fioc97}. This sets the emission rate of ionizing photons per unit stellar mass as $2.85\times 10^{45} ~s^{-1} M_{\odot}^{-1}$. In addition to the stellar contribution, we also consider an X-ray spectrum as a power-law of the energy with a spectral index fixed to $1.2$. The emission rate of the X-ray photons per unit stellar mass is set as $f_{\rm X} \times 10^{42} ~s^{-1} M_{\odot}^{-1}$ where we vary the X-ray heating efficiency $f_{\rm X}$. We define the UV band as an energy range of 13.6 eV to 100 eV, while the X-ray band span from 100 eV to 10 keV. Such SEDs are also used in our previous studies, such as \cite{ghara15, 2019MNRAS.487.2785I, ghara20}. The source model used in this paper also assumes that only those dark matter halos with masses larger than $M_{\rm h,\, min}$ emit Ly$\alpha$, UV and X-ray photons. We also vary $M_{\rm h,\, min}$ in this study. Table \ref{table1} shows the four different values for each of $f_{\rm X}$ and $M_{\rm h,\, min}$ as considered in this study. We consider $16$ different Cosmic Dawn scenarios, which correspond to different combinations of $f_{\rm X}$ and $M_{\rm h,\, min}$.

Note that our large simulation volume does not consider low mass halos ($\mh < 10^9 M_{\odot}$) because resolving these halos in this large volume requires simulations of a very high dynamic range which are beyond the available computing resources. Even if we include the mini-halos using an approximation technique, e.g., a subgrid model \cite{barkana04,ghara15a,2020MNRAS.494.3294N}, the overall conclusions regarding the bispectrum are expected to remain valid (for details, see section 3.2 of \cite{ghara15a}, which shows the importance of mini-halos in the context of the power spectrum).

\section{Bispectrum estimation}
\label{sec:bis}


In this section, we discuss the formalism for estimating the bispectra for all possible unique triangles in Fourier space.

\subsection{The bispectrum and the different triangle configurations}
\label{sec:trig_config}

We use the definition of the bispectrum estimator given in \cite{majumdar18} to estimate the bispectra from the simulated data. The bispectrum estimator for the $i^{\rm th}$ triangle configuration bin is given as
\begin{equation}
  \hat{B}_{i}({\bf k}_1, {\bf k}_2, {\bf k}_3) = \frac{1}{N_{{\rm tri}}V} \sum_{
[{\bf k}_1 +{\bf k}_2 +{\bf k}_3 =0] \in i }\Delta_{\rm b}(\textbf{k}_1) \Delta_{\rm b}(\textbf{k}_2) \Delta_{\rm b}(\textbf{k}_3) \, ,
\label{eq:bispec_est}
\end{equation}
where $\Delta_{\rm b}(\textbf{k})$ is the Fourier transform of the differential brightness temperature of the 21-cm field, $N_{\rm tri}$ is the number of samples of closed triangles associated with the $i^{\rm th}$ triangle configuration bin, $V$ is the simulation volume. The actual calculation of the bispectrum has been performed using the algorithm presented in \cite{majumdar18} and \cite{majumdar20}. This algorithm has introduced the following constraints in order to define the shapes of the $k$-triangles, which further reduces the computation time for bispectrum estimation,
\begin{equation}
   n = \frac{k_2}{k_1} 
    \label{ratio}
\end{equation}

\begin{equation}
    \cos{\theta} = -\frac{{\mathbf k}_1\cdot {\mathbf k}_2}{k_1 k_2},
    \label{cos_theta}
\end{equation}
where $k_1$ and $k_2$ are the magnitudes of the vectors ${\mathbf k}_1$ and ${\mathbf k}_2$ respectively.

 \subsection{The unique triangle configurations in the triangle parameter space}
 \label{sec:uni_trig_config}

For a given size of $k$-triangle, which is determined by the magnitude of the $k$ modes involved, its shape can be uniquely specified in the $n-\cos\theta$ space by imposing the conditions prescribed in \cite{bharadwaj20}, i.e.,
 \begin{eqnarray}
&&{k}_1 \geq {k}_2 \geq {k}_3
\label{eq:k_relation}\\
&&0.5 \leq n \leq 1.0
\label{eq:n}\\
&&0.5 \leq \cos\theta \leq 1.0.
\label{eq:costheta}
\end{eqnarray}

The unique triangles are indicated by the shaded region in the $n-\cos{\theta}$ space where $n\cos{\theta} \geq{0.5}$ as shown in the right panel of Figure 1 in \cite{majumdar20}. The shaded region in the left panel of this figure is formed by all possible points of intersection between the $\textbf{k}_2$ and $\textbf{k}_3$ vectors while equations \eqref{eq:k_relation}, \eqref{eq:n} and \eqref{eq:costheta} are satisfied. The left panel of this figure also shows all possible unique $k$-triangles represented by the cosine of the angle $\chi$ subtended by the vertex of triangle facing the $k_1$ arm. For a detailed classification of unique $k$-triangles we refer the reader to section 2.2 of \cite{majumdar20}.

We estimate and analyze the spherically averaged bispectra extracted from the data cubes simulated at the target redshifts during CD. For this, we divide the entire $n$ as well as $\cos{\theta}$ range with step sizes $\Delta n=0.05$ and $\Delta\cos{\theta}=0.01$. We further bin the entire $k_1$-range ($k_{\rm min} = 2\pi$/[box size] $\approx 0.01 \mp$ , $k_{\rm max} = 2\pi$/2[grid spacing]$ \approx 2.64 \mp$) into $15$ logarithmic bins. We label each bin by the value of $k_1$ which determines the size of the $k$-triangle in our formalism.

\section{Results}
\label{sec:results}
\begin{figure*}
  \includegraphics[width=1\textwidth,angle=0]{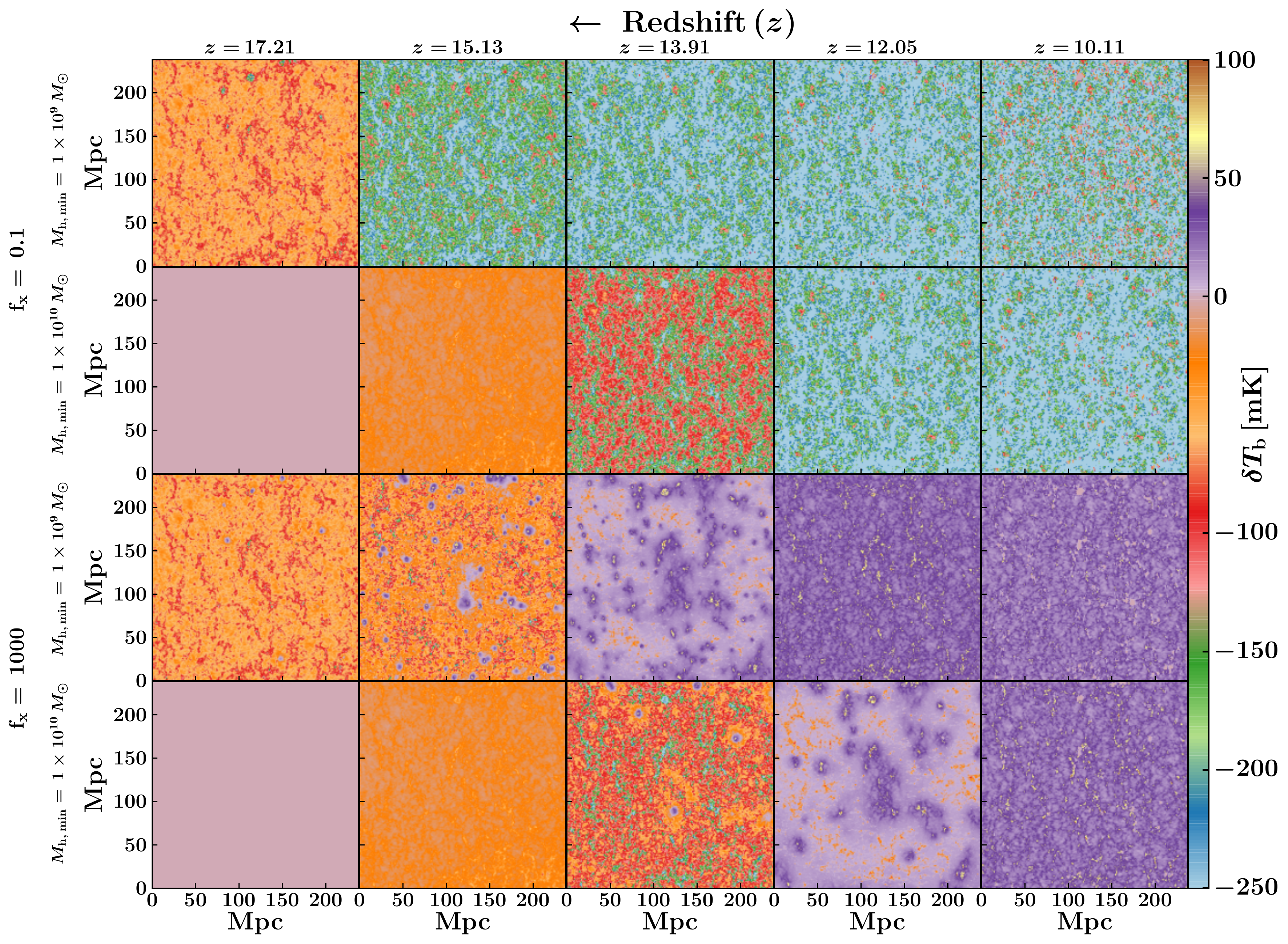}
  \caption{Zoomed-in slices (side length $238.09$ Mpc, out of the full  $(714.29 {\, \rm Mpc})^3$ box size) of the differential brightness temperature ($\TB$) for four different CD scenarios showing the redshift evolution of the 21-cm signal. The upper two panels show the redshift evolution of $\TB$, where $\fx = 0.1$ (smallest), and $\mh$ attains two extreme values of the considered range. The lower two rows of panels have $\fx = 1000$ (largest) with the same $\mh$ values as the upper panels.}
  \label{fig:Tb_map}
\end{figure*}
In this section, we present the analysis of the 21-cm bispectrum from the CD for several scenarios corresponding to the different combinations of the source parameters, as listed in Table \ref{table1}. Fig. \ref{fig:Tb_map} shows the slices of $\TB$ maps for four different scenarios with different ($\mh$, $\fx$) pairs formed by the combinations of two extreme values of these parameters. Here panels in each row show the redshift evolution of a specific CD scenario with decreasing redshifts from left to right. The top two rows of panels show the redshift evolution of $\TB$, where $\fx = 0.1$ (smallest) and $\mh$ is kept fixed at the two extreme values of its range listed in Table \ref{table1}. Similarly, the bottom two rows of panels have $\fx = 1000$ (largest) with the same $\mh$ values as the two top rows of panels. Figure \ref{fig:Tb_map} shows that signal strength is low at $z \sim 17$ due to inefficient Ly$\alpha$ coupling in any CD scenario. At $z \sim 14$ the maps for $\fx = 0.1$ and $1000$ are different as emission regions expand very fast for $\fx=1000$ scenarios. Furthermore, the scenarios with $\fx = 0.1$ remain absorption signal-dominated throughout the redshift range considered here. However, the signal for $\fx = 1000$ scenarios already becomes emission dominated at the late stages of the CD.

\begin{figure*}
  \includegraphics[width=1\textwidth,angle=0]{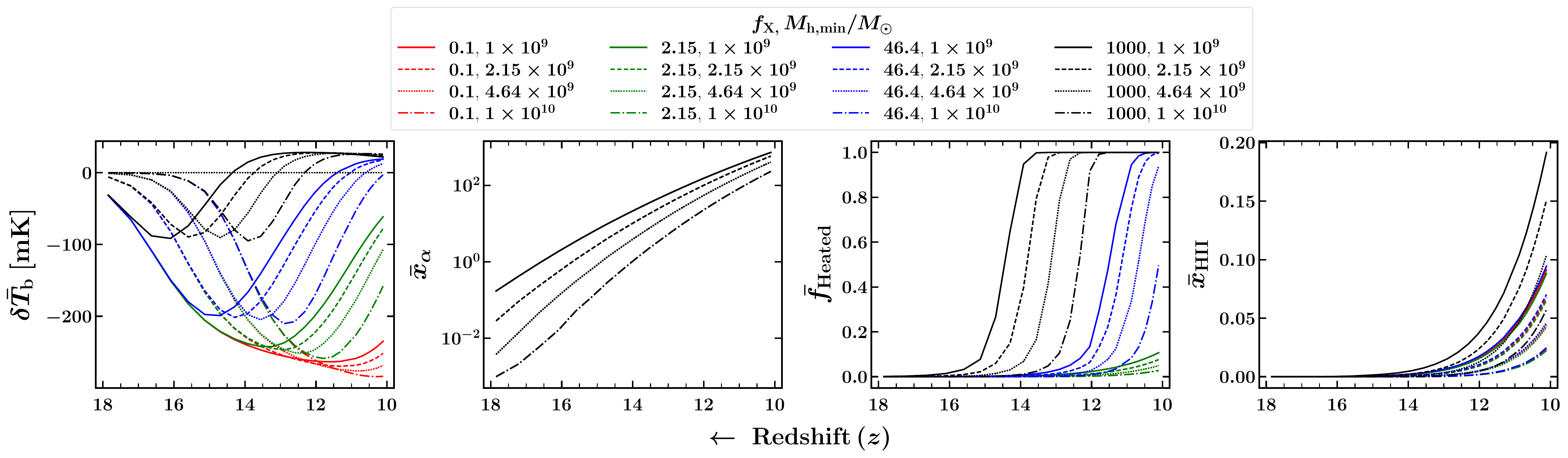}
  \caption{Shown are the global evolution of:
  (left) the mean 21-cm differential brightness temperature ($\tb$) for all CD scenarios; (middle left) the mean Ly$\alpha$ coupling coefficient; (middle right) the fraction of the simulation volume heated above the CMB temperature (i.e. $T_{\rm g} > T_{\rm CMB}$); and (right) the mean ionization fraction ($\xhii$).}
  \label{fig:Tb_z}
\end{figure*}

In Fig.~\ref{fig:Tb_z}, the left panel shows the redshift evolution of the global 21-cm brightness temperature ($\tb$) for all CD scenarios considered here. One can see that the amplitude of the absorption trough remains small for larger values of $\fx$. The transition from absorption signal to emission happens earlier for larger $\fx$ values. The middle panels show the redshift evolutions of the mean Ly$\alpha$ coupling coefficient ($\bar{x}_{\alpha}$) and the fraction of the simulation volume heated above the CMB temperature ($\bar{f}_{\rm Heated}$), i.e., $T_{\rm g} > T_{\rm CMB}$, respectively. The Ly$\alpha$ coupling coefficient is independent of the $\fx$ values and varies only with the $\mh$ values. Therefore, for smaller $\mh$ the values of $\bar{x}_{\alpha}$ are larger. The $\bar{f}_{\rm Heated}$ will depend on both $\fx$ and $\mh$ values. For example, the heating for scenarios with higher $\fx$ and lower $\mh$ values will begin and saturate earlier. The right-most panel in Figure \ref{fig:Tb_z} shows the redshift evolution of the global ionization fraction ($\xhii$) for all CD scenarios. In this study, we assume that X-ray photons will also engage in a small amount of photo-ionization via their X-ray photons on top of the usual dominant UV photo-ionization. Therefore, photo-ionization will depend on both $\mh$ and $\fx$ values. The reionization will proceed faster in the scenarios with smaller $\mh$ and higher $\fx$ values. Thus it is clear that the Ly$\alpha$ coupling, thermal, and ionization histories of the IGM are different in the different CD scenarios due to their strong dependence on the CD source properties and thus result in different CD histories ($\tb$ vs $z$) in different scenarios. 

In this work, we assume that the star-forming galaxies are the dominant sources of ultra-violet (UV) photons, and mini-QSOs are the only sources of X-ray photons. Both of these sources are located in halos above a certain minimum mass. This implies that all the astrophysical processes, such as the Ly$\alpha$ coupling, X-ray heating, and UV ionization, vary with $\mh$ (see Figure \ref{fig:Tb_z}). We further consider different $\fx$ values for each $\mh$. The $\fx$ values will determine the X-ray photon production rate, with low values corresponding to faint and large values to bright sources.
These sources will yield different IGM heating signatures (see the third panel of Figure \ref{fig:Tb_z}). One important point to note here is that for each CD scenario considered here, all of the possible astrophysical processes are running simultaneously (in a competitive manner with each other) from the beginning of the CD. For a particular CD scenario, which astrophysical process dominates over the rest at a specific stage of the CD compared to other scenarios, is solely dependent on the combination of the $\fx$ and $\mh$ values considered for that scenario.

For all the simulated CD scenarios, generated using all possible combinations of parameter values listed in Table \ref{table1}, we have estimated the 21-cm power spectrum [$P(k)$] and bispectrum [$B(k_1, k_2, k_3)$]. We estimate the bispectrum as discussed in \citep{majumdar18}. Further, we follow the convention of \citep{kamran20} and \citep{majumdar20} to demonstrate our result in terms of spherically averaged normalized power spectrum and bispectrum that are respectively defined as $\Delta^2(k) = \big[ k^3 P(k)/(2\pi^2)\big]$ and $\Delta^3(k_1, n, \cos{\theta}) = \big[ k_1^3 k_2^3 B(k_1, k_2, k_3)/(2\pi^2)^2\big]$. To get the better insights of the signature of the sources with different properties on the IGM, we first discuss the evolution of $\Delta^2(k)$ with redshifts, before discussing the relation between $\Delta^3(k_1, n, \cos{\theta})$ and different source signatures.

\subsection{The redshift evolution of the power spectrum}

\begin{figure*}
  \includegraphics[width=1\textwidth,angle=0]{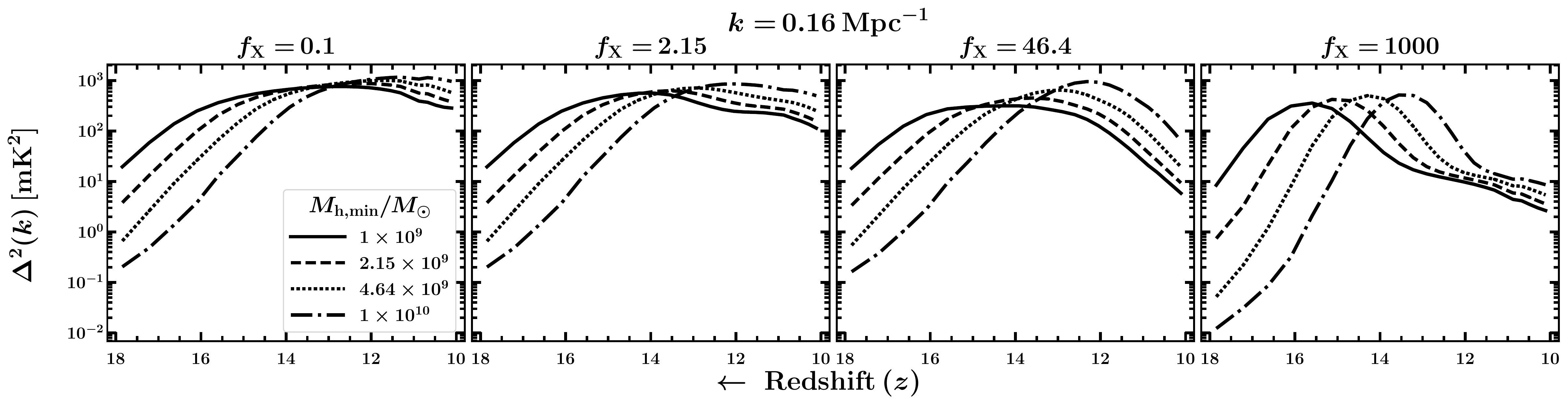}
  \caption{The redshift evolution of the power spectra at $k = 0.16 \mp$. Each panel corresponds to a fixed $\fx$, with lines for the different $\mh$ values, as labelled.}
  \label{fig:PS_z}
\end{figure*}

In Figure~\ref{fig:PS_z} we show the redshift evolution of the power spectra [$\Delta^2(k)$] at $k = 0.16 \mp$. Each panel represents the evolution of the same for a fixed $\fx$ and all possible $\mh$ values. It is apparent that $\Delta^2(k)$ at a particular $\fx$ strongly depends on the value of $\mh$. At the early stages of the CD, the power spectra for the low $\mh$ sources have a higher magnitude than that for high $\mh$ sources. This is because, in the case of the low $\mh$ sources, very low mass halos ($M_{\rm h} = 10^{9}$--$10^{10}\, M_{\odot}$), which are numerous in number compared to the high mass halos, are also contributing. Hence, these sources will produce a relatively larger amplitude of fluctuations in the signal via a comparatively stronger Ly$\alpha$ coupling process compared to the case when these low mass sources are ignored (see two left panels of Figure \ref{fig:Tb_map} or second panel of Figure \ref{fig:Tb_z}). During the late stages of the CD, the heating becomes more prominent for low $\mh$ sources together with the photo-ionization. This is because in this case, very low mass halos ($M_{\rm h} = 10^{9}$--$10^{10}\, M_{\odot}$), which are numerous in number are also contributing. Therefore the efficient heating by low $\mh$ sources decreases the amplitude of fluctuations in the 21-cm signal during the late stages of the CD at a faster rate compared to high $\mh$ sources. This results in the low $\Delta^2(k)$ for the low $\mh$ sources scenario. Thus, at the late stages of the CD, $\Delta^2(k)$ lower for low-$\mh$ cases compared to high-$\mh$ ones. For the sources at a fixed $\fx$ and $\mh$, $\Delta^2(k)$ first increases, reaches the maxima, and then decreases. The steepness with which  $\Delta^2(k)$ decreases solely depends on how prominent the IGM heating is by that time. Thus, the larger the rate of heating, the steeper the slope of decrease (see Figure \ref{fig:PS_z} as we move from left to right panels). One can pose the question, what determines the redshift at which $\Delta^2(k)$ starts to decline? This decline in  $\Delta^2(k)$ happens when the gas heating is more important than is the Ly$\alpha$ coupling. In principle, this occurs at the time when the emission (heated) regions in the IGM grow and the absorption decreases. However, this does not mean the heating was inefficient before this redshift. Heating actually became important at a somewhat higher redshift than this. This phenomenon, in principle, makes the IGM warm (or less cold) by decreasing the amplitude of the absorption signal around the sources before producing the heated regions around the same sources. This duration depends strongly on the source parameters and is probed directly by the bispectrum, but is only weakly reflected in the power spectrum. In the following sections, we explore to what extent bispectrum can robustly distinguish the various source models.

\begin{figure*}
  \includegraphics[width=1\textwidth,angle=0]{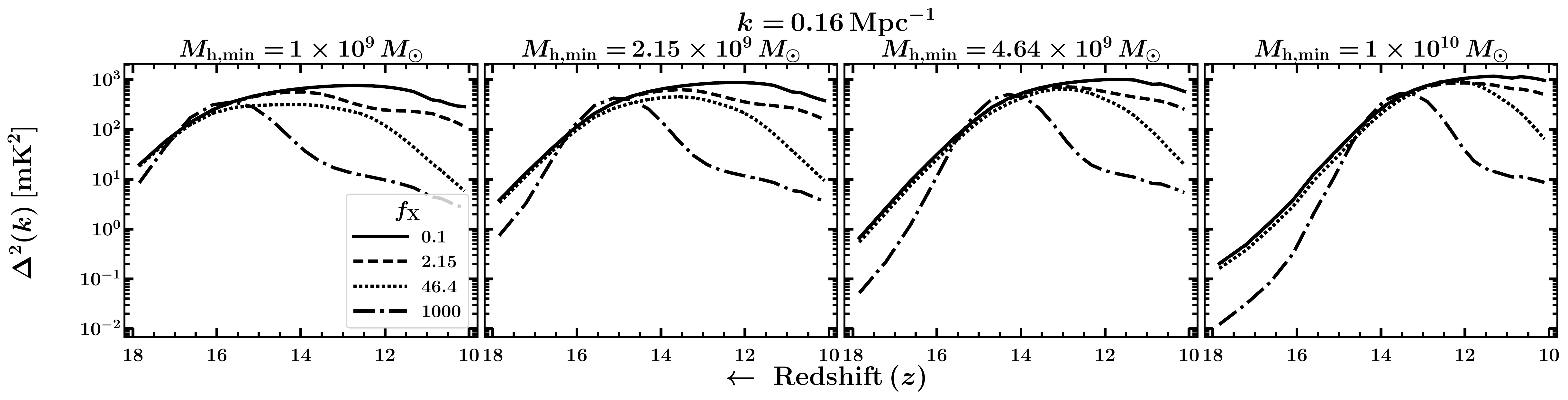}
  \caption{The redshift evolution of the power spectra at $k = 0.16 \mp$. Each panel corresponds to a different $\mh$ value, with lines for the different $\fx$ values, as labelled.}
  \label{fig:PS_z1}
\end{figure*}

In Figure \ref{fig:PS_z1} we show the redshift evolution of the power spectra with varying $\fx$ values for fixed values of $\mh$. The one generic feature of the large scale power spectrum is that for $\fx$ values in the range $0.1-100$ the power spectrum remains almost same in amplitude at high redshifts ($\sim 18-15$). Only for extremely high values of $\fx \sim 1000$, the amplitude of power spectrum shows a significantly different evolution. Furthermore, as we move towards higher $\mh$ values (Figure~\ref{fig:PS_z1}, right), the differences between the power spectra for different $\fx$ values become ever less significant. These results are consistent with our earlier discussion related to Figure \ref{fig:PS_z}.

\subsection{The redshift evolution of the bispectrum}
\begin{figure*}
  \includegraphics[width=1\textwidth,angle=0]{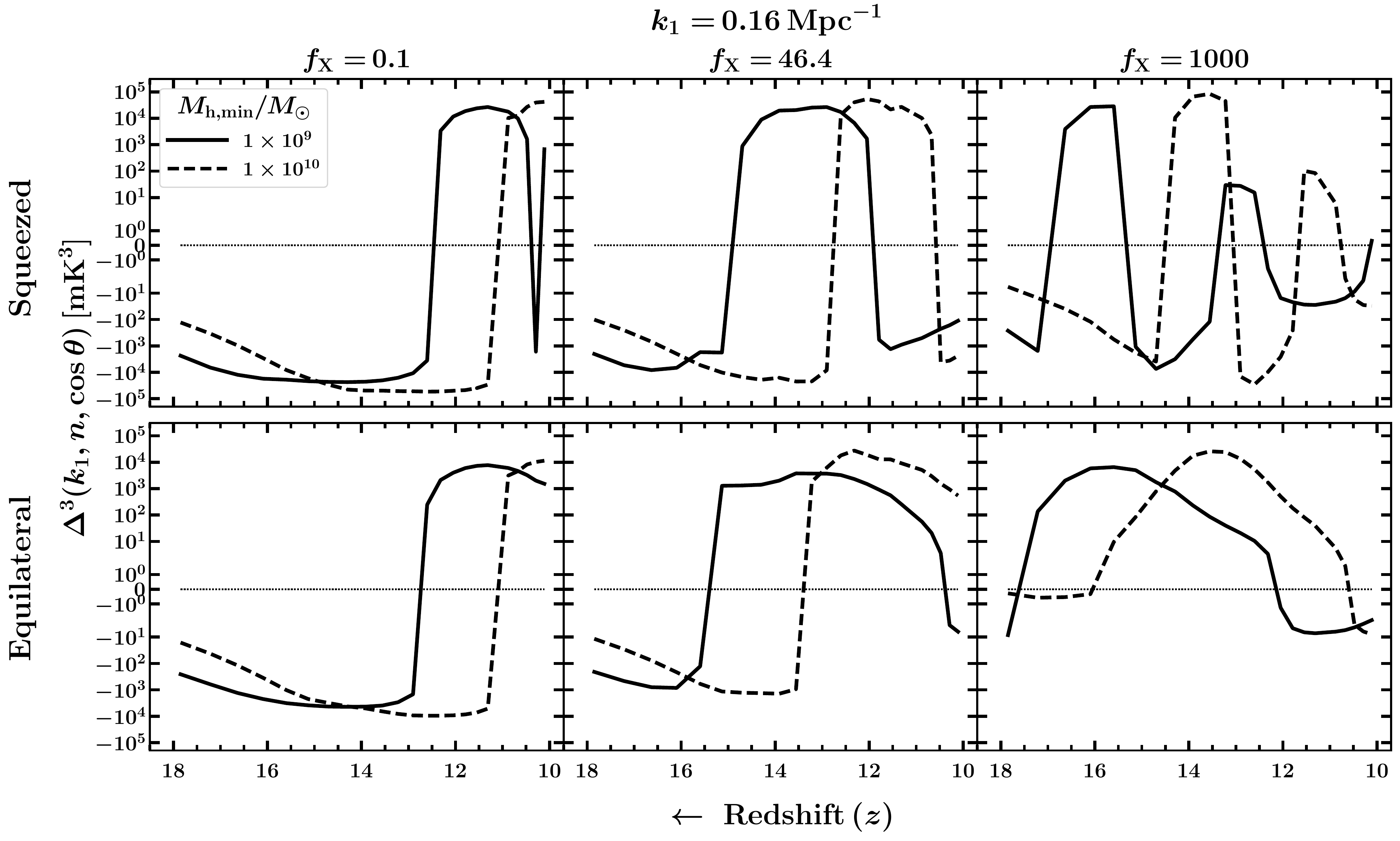}
  \caption{The evolution of the bispectra at $k_1 = 0.16 \mp$. in (top panels) the squeezed limit; and (bottom panels) 
  equilateral $k$-triangles. For each panel the bispectra are plotted at a fixed $\fx$, but with two extreme values of $\mh$.}
  \label{fig:BS_z_eq_sq}
\end{figure*}

\subsubsection{Squeezed-limit and equilateral bispectrum}
\label{sec:sq_eq_bs}
In Fig.~\ref{fig:BS_z_eq_sq} we show the redshift evolution of the bispectra at $k_1 = 0.16 \mp$ for fixed $\fx$ and two extreme values of $\mh$ in the squeezed limit (top panels) and for equilateral $k$-triangles (bottom). At first glance, one can see that irrespective of the specific triangle configurations, the shapes of the bispectra for a fixed value of $\fx$ appear to be the same for both $\mh$ values. The difference between the features of the bispectra in these two scenarios is that they appear at the different redshifts. Thus, the feature in the signal bispectrum for low $\mh$ sources is shifted towards higher redshifts compared to high $\mh$ sources. We also find similar power spectra shapes. The physical reason behind it is that for low-$\mh$ scenarios, the low-mass sources ($< 10^{10}\, M_{\odot}$) are present and  begin to impact the IGM through the aforementioned astrophysical processes earlier compared to the high $\mh$ scenarios. The dominant IGM processes are shared for both low- and high-$\mh$ scenarios, resulting in topologically-similar distributions for the IGM 21-cm signal. This yields similar shapes or features in any of the statistical measures of the 21-cm signal, only shifted by the redshifts at which these features appear. This then raises the question: if different $\mh$ scenarios result only in shifting of the features in the bispectra with redshift (similar to the behaviour observed in the power spectra too), then how can the bispectrum be a better statistic for distinguishing these source models? The reason is that these source models, depending on their respective source parameter values, invoke different levels of the non-Gaussianity in the signal, the time evolution of which can be tracked through the evolution of both bispectrum magnitude and sign. The sign of the bispectrum and its time evolution is a unique tracer of all the dominant physical processes in the IGM, which are determined by the source parameter values (for details, we refer the interested reader to \cite{kamran21}).

Next we discuss how the bispectrum evolves with redshift for fixed $\fx$ and $\mh$ (see Fig.~\ref{fig:BS_z_eq_sq}). We first focus on the evolution of the bispectrum for squeezed limit $k$-triangles and then compare it to the case of equilateral $k$-triangles to see which might be a more suitable probe of the different source models. For $\fx = 0.1$, from the very early stages of the CD, the magnitude of bispectrum for squeezed limit $k$-triangle increases. This increment lasts while the $\mh$ source model dominates the 21-cm signal fluctuations via the Ly$\alpha$ coupling process. For instance, this redshift for $\mh = 1 \times 10^9\, M_{\odot}$ sources is $z \sim 13.5$. Although the heating is too slow for $\fx = 0.1$, the bispectrum is highly sensitive to the heating process. Thus when eventually heating becomes dominant over Ly$\alpha$ coupling, the increase in bispectrum magnitude stops. These processes impact the large-scale fluctuations in the 21-cm signal; thus, the small $k$ squeezed-limit signal bispectrum is more sensitive to them. The heating first reduces the amplitude of the absorption signal around the sources by producing warm (or less cold) regions and then produces the heated (emission) regions around the same sources. The reduction in the magnitude of the bispectrum in the redshift range $13.5 \gtrsim z \gtrsim 12.5$ is the signature of this phenomenon. Once a significant fraction of the IGM is heated, this will cause the bispectrum to become positive (i.e. a sign change in the bispectrum). The bispectrum remains positive until another physical process (i.e. photo-ionization) becomes dominant over the heating. As ionization becomes important, there will be a race between the heating and ionization processes at late CD stages for $\fx = 0.1$ sources. The ionized regions will form in the already heated regions. The heated regions, by this time, start to act as a large-scale background for the ionized regions. Previously, \cite{bharadwaj05a} has shown that the Fourier equivalent of the 21-cm fluctuations (i.e. $\Delta_{\rm b}$) at a large length scale due to the ionization in heated background will be negative. This will result in a negative bispectrum at late CD stages (i.e. another sign change). However, even at late stages for $\fx = 0.1$ scenarios, all of the heated regions have not merged together to form a uniformly heated background. Further, by this time, the ionization rate is higher than the heating rate (see third and fourth panels of Figure \ref{fig:Tb_z}), which eventually results in the ionization front surpassing the heating front. Once this transition happens, the ionized regions now appear in a Ly$\alpha$ coupled background (see the last panel of Figure \ref{fig:Tb_map} for $\fx = 0.1$, $\mh = 1\times10^9\, M_{\odot}$). In \cite{kamran21}, we have shown that the large scale $\Delta_{\rm b}$ dominated by the ionized regions in a Ly$\alpha$ coupled IGM will be positive. This will correspond to the positive bispectrum at the very late stage of the CD (another sign change).

Next, we discuss how the squeezed-limit bispectra evolve with redshifts for different $\fx$ values, irrespective of the value of $\mh$. As we move from left to right panels in Figure \ref{fig:BS_z_eq_sq}, the $\fx$ values increases logarithmically. The feature of the sign change of the squeezed limit bispectra that we discussed for $\fx = 0.1$ also remains for higher $\fx$ values. The only difference is that this feature appears at earlier redshifts for higher values of $\fx$. In addition to this, we observe a follow-up sign change from positive to negative. This second sign reversal is a signature that all the heated regions are being connected to form a heated background. On this background, a few leftover cold absorbing regions are embedded, and their fluctuations dictate the 21-cm fluctuations. This results in a negative squeezed-limit bispectrum. Due to further heating, the leftover absorbing regions also diminish in volume to make the entire IGM uniformly hot. This causes the magnitude of the negative bispectra to decrease with the decreasing redshifts. To see the connection of bispectrum with the IGM physics in detail, we refer the reader to our recent work \citep{kamran21}.

Interestingly, for $\fx = 1000$ at later times we observe two additional sign changes in the bispectra (designated as third and fourth sign changes, respectively). We first interpret the third sign change by considering the source model with $\mh = 1 \times 10^9\, M_{\odot}$ for instance (solid black line). For $\fx =1000$ the heating of the IGM proceeds much faster than for lower $\fx$ values (middle right in Figure \ref{fig:Tb_z}). Further heating thus will not contribute to the 21-cm fluctuations. The photo-ionization in this scenario is also not dominant by the redshift $z \sim 13.5$ (see the solid black line in the right-most panel of Figure \ref{fig:Tb_z}) at which the sign change is being observed. Thus the fluctuations in the IGM 21-cm signal can be dominated by the matter density fluctuations alone. It has been well established that the bispectrum sign will be positive if the matter density fluctuations dictate the 21-cm fluctuations in heated IGM (\cite{majumdar18,majumdar20}). That is why we observe a positive bispectrum in the redshift range $13.5 \gtrsim z \gtrsim 12.5$. Around $z \sim 12.5$, the photo-ionization becomes the dominant process in an already heated IGM. Previously, \cite{majumdar18,majumdar20} have also shown that the bispectrum sign due to the neutral fraction fluctuations in a heated IGM will be negative. For $z \lesssim 12.5$, therefore, we observe a negative bispectrum (the fourth sign change of the bispectrum).

Finally, we compare the evolution of the bispectra for equilateral $k$-triangles vs. the squeezed-limit $k$-triangles. In Figure \ref{fig:BS_z_eq_sq} (bottom panels) we show the redshift evolution of the bispectra for equilateral $k$-triangles. The features in the shapes of the equilateral bispectra for $\fx = 0.1$ and $46.6$ match with the ones observed in the squeezed limit. This implies that the equilateral bispectra are able to probe the same signal characteristics that the squeezed-limit bispectra are probing. The bispectra for these two triangle shapes only differ by the redshift values at which their respective features appear. The reason for this is the following: for $k_1 = 0.16 \mp$ (designated as the ``large scale''), the squeezed-limit triangle satisfies the condition $k_1 = k_2 = 0.16 \mp > k_3 \rightarrow 0$. The bispectrum for this triangle shape provides the correlation between the signal fluctuations at two different length scales i.e. large and very large scales. On the other hand, the equilateral bispectrum for $k_1 = k_2 = k_3 = 0.16 \mp$ provide the correlation between signal fluctuations at the same length scales i.e. large scale. Thus, the ways in which the bispectra for these two triangle shapes behave in order to characterize the signal are different. For the same reason, the features of the equilateral bispectra for $\fx = 1000$ significantly differ from that of the squeezed one. Unlike the squeezed limit bispectra, where its evolution features a total of four sign changes during the entirety of the CD, the equilateral bispectra show only two sign changes. These two sign changes are far apart in redshifts compared to the first two sign changes in the squeezed-limit bispectra. The first sign change from negative to positive appears at a very early stage, whereas the second one from positive to negative appears at a very late stage of the CD. The equilateral bispectra thus remain positive for most of the time during the CD. Hence, the equilateral bispectrum is able to follow the evolution of the heated regions in the IGM for a long duration compared to the squeezed-limit bispectrum.

\subsubsection{L-Isosceles bispectrum}
\label{sec:liso_bs}
\begin{figure*}
  \includegraphics[width=1\textwidth,angle=0]{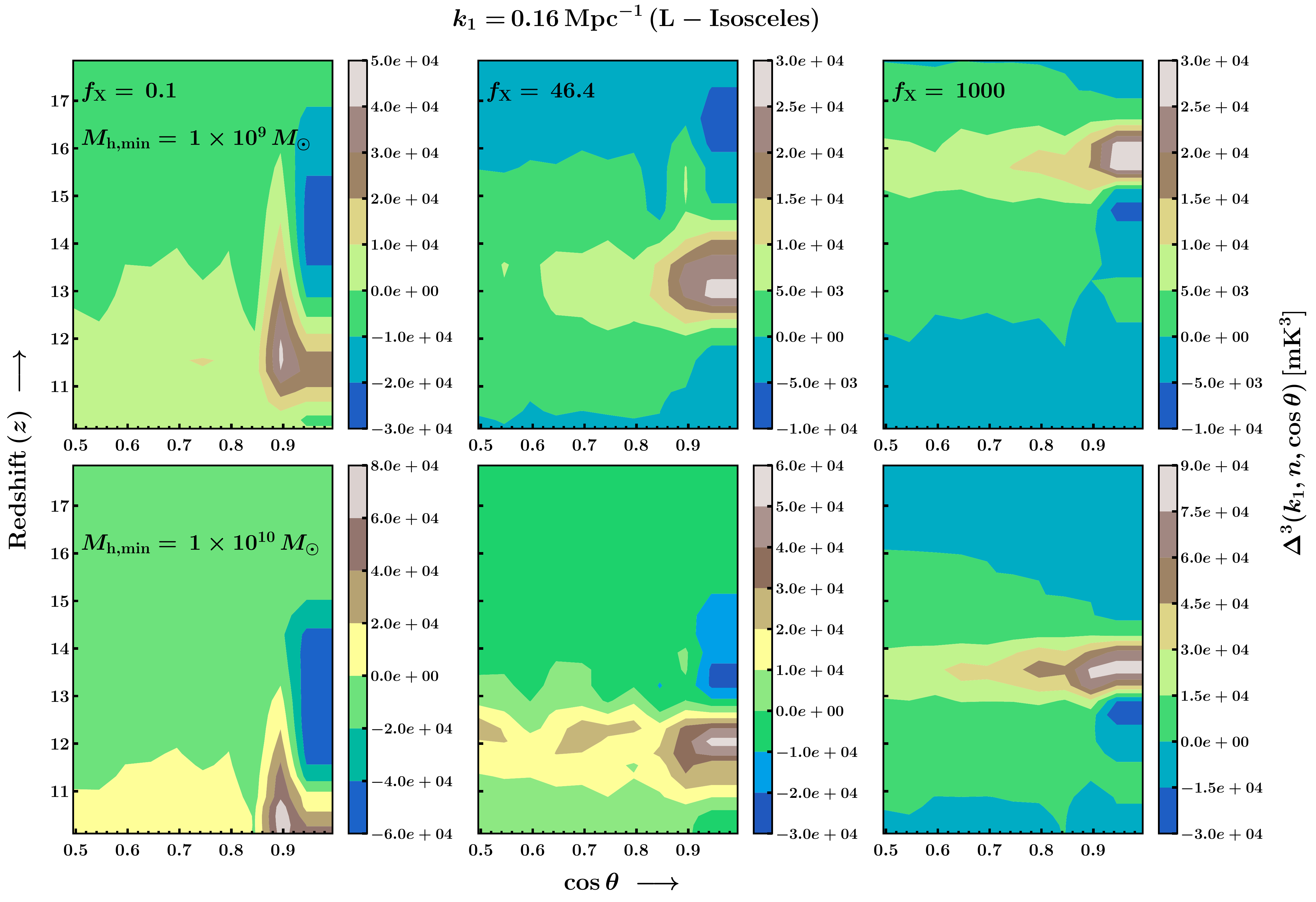}
  \caption{Evolution of the L-isosceles bispectra for $k_i=0.16\,Mpc^{-1}$ for three different $\fx$ values (as labelled) for (top) $\mh = 1\times10^9\, M_{\odot}$; and (bottom) for $\mh = 1\times10^{10}\, M_{\odot}$. }
  \label{fig:cost_z_bs_liso}
\end{figure*}
The squeezed and equilateral triangle shapes are special cases of a more generic triangle shape i.e. L-isosceles (L refers to `Large') triangles. It is therefore useful check how well the L-isosceles bispectra retain the features of the signal as probed by the squeezed and equilateral bispectra.  L-isosceles are defined as  $k$-triangles that satisfy the condition $k_1 = k_2 \geq k_3$, i.e. two large arms of each triangle are equal. In the $n$--$\cos{\theta}$ space (shaded region in the right panel of Figure 1 in \cite{majumdar20}), the horizontal line with $n = 1$ and $\cos{\theta} \in [0.5,1]$ is the representation of the L-isosceles $k$-triangles.  The two endpoints on the L-Isosceles line are the equilateral ($\cos{\theta} = 0.5$) and squeezed ($\cos{\theta} = 1.0$) $k$-triangles. In order to get an understanding of about up to what extent the observed features of the squeezed-limit and equilateral signal bispectra from CD could be generalized, we estimate the bispectra for L-Isosceles $k$-triangles at $k_1 = 0.16 \mp$. In Figure \ref{fig:cost_z_bs_liso} (top panels) show the redshift evolution of the L-Isosceles bispectra for three different $\fx$ values while keeping $\mh$ fixed ($1\times10^9\, M_{\odot}$). The bottom panels show the same at $\mh = 1\times10^{10}\, M_{\odot}$. For a fixed $\fx$, the effect of $\mh$ on the L-isosceles bispectra is the same as it is on the squeezed-limit and equilateral bispectra. For instance, the shapes of the bispectra for different $\mh$ are the same. They only differ by the redshift values at which their features appear. However, the magnitude of the bispectra might get altered with changing $\mh$. This depends on how these different $\mh$ sources impact the IGM. Further, for a fixed $\mh$, the effect of varying $\fx$ that we have observed already in the case of the squeezed-limit bispectrum redshift evolution, extends to the L-isosceles which are in the vicinity of squeezed $k$-triangle bispectra i.e. with $\cos{\theta} \in (0.8,1)$. Under the same conditions, the features in the equilateral bispectrum evolution can also be seen to be propagating to L-isosceles bispectra which are within its vicinity i.e. $\cos{\theta} \in (0.5,0.8)$. This can be clearly understood through the following example: for sources with $\mh = 1\times10^9\, M_{\odot}$, the effect of X-ray heating with $\fx = 1000$ causes the bispectra for squeezed limit and triangle shapes in its vicinity to change its sign four times. Similarly, the bispectra for equilateral and triangle shapes in its vicinity change their sign only twice.

\subsection{Evolution of the bispectrum in the source parameter space}
\begin{figure*}
  \includegraphics[width=1\textwidth,angle=0]{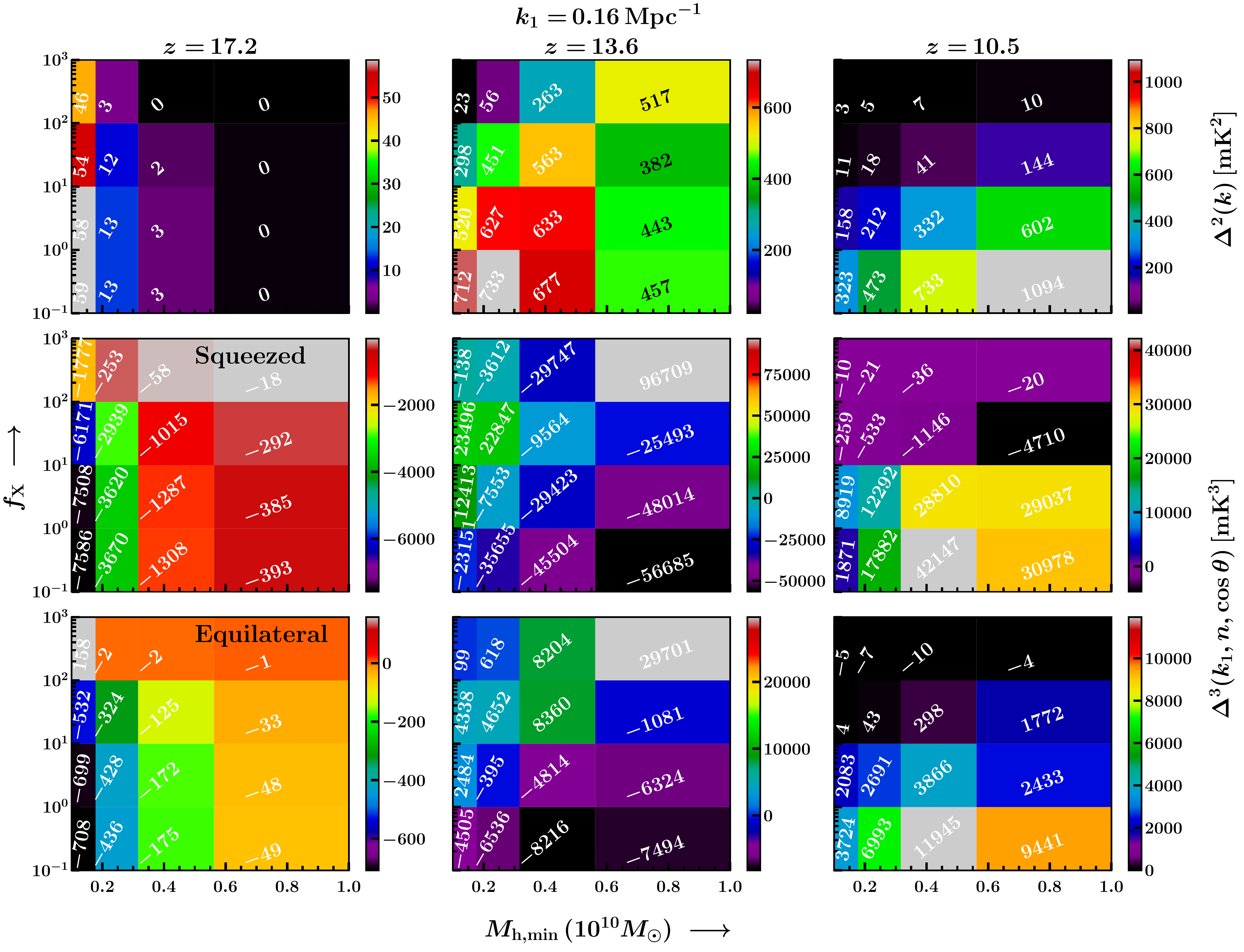}
  \caption{The evolution in phase-space formed by $\mh$ and $\fx$ values for $k_1 = 0.16 \mp$ of: (top) power spectra;
  (middle) bispectrum in the squeezed-limit $k$-triangles; 
  and (bottom) bispectrum for equilateral $k$-triangles.
  Each column of panels is at three different redshifts, as labelled, designated as {\em early}, {\em intermediate}, and {\em late} CD stages respectively from left to right.} 
  \label{fig:Mmin_fx_BS}
\end{figure*}
In this section, we present a comparative study of the power spectrum and bispectrum evolution in the parameter space of $\fx$ and $\mh$. Figure \ref{fig:Mmin_fx_BS} shows the evolution of the power spectrum (top panels) and bispectrum for squeezed (middle) and equilateral $k$-triangles (bottom) for $k_1 = 0.16 \mp$. Each case is shown at three different stages of the CD: {\em early} stages of the CD, where the Ly$\alpha$ coupling process dictates the 21-cm fluctuations in the cold IGM (Fig.~\ref{fig:Mmin_fx_BS}, left column); {\em intermediate} stage where the X-ray heating competes with the Ly$\alpha$ coupling (Fig.~\ref{fig:Mmin_fx_BS}, middle column), where depending on the CD scenarios, this competition can be weaker or stronger; and {\em late} stages of the CD, where the Ly$\alpha$ process for all CD scenarios and X-ray heating for the scenarios with the highest $\fx$ values are saturated
(Fig.~\ref{fig:Mmin_fx_BS}, right column). The ionization thus becomes important after heating in the case of this highest $\fx$ scenario. The number on each pixel is an estimate of the respective signal statistic (power spectrum or bispectrum) corresponding to that pixel. A visual inspection of this figure reveals that the evolution of the power spectrum magnitude is very slow compared to the variation in magnitude of the bispectrum. On top of this, an additional advantage of the bispectrum is that its sign depends on varying source parameter values, thus providing additional information. The magnitude and sign evolution of the bispectrum in this parameter space makes it a more robust probe of the different source models.

Next, we explore how the bispectrum changes its shape, magnitude and sign with varying source parameters and thus potentially provides a unique and robust probe of the source models. For this, we focus on how the features in the bispectrum capture the changes in the IGM physical processes arising due to the variations in the source parameters. At the {\em early} stage of the CD, the squeezed-limit bispectra are negative in the entire phase space. In our earlier work \citep{kamran21}, we have shown that during the {\em early} stages of the CD, the bispectrum will be negative when the Ly$\alpha$ coupling dictates the 21-cm signal fluctuations in the uniformly cold/heated background IGM. It is, therefore, evident that whatever be the source models, if the dominant process is the Ly$\alpha$ coupling in the uniformly cold/heated IGM, then the bispectrum will be negative. The equilateral bispectra are also negative for most of the parameter space except for the scenario having the lowest $\mh$ and highest $\fx$ values, in which the bispectrum is positive. Previously, \citep{kamran21} has also established that during the CD, the bispectrum will be positive when the signal is coming from the emitting (heated) regions placed in a cold background. This positive equilateral bispectrum is the signature that the fluctuations induced by the heated regions generated for this specific source model are being probed. This comparative analysis confirms that the equilateral bispectrum can probe even smaller heated regions formed at the beginning of the CD compared to the squeezed-limit bispectrum. This is because the squeezed-limit bispectrum probes the comparatively larger-scale features than the equilateral bispectrum.

Furthermore, at the {\em early} stage, the region in the parameter space where the magnitude of the negative bispectrum peaks, corresponds to strong Ly$\alpha$ coupling. The magnitude of this negative bispectrum decreases with increasing $\fx$ (from bottom to top) while $\mh$ stays constant. We explain this phenomenon as follows. The increasing $\fx$ values result in an increased rate of heating. The heating around the sources, as it progresses with time, has two significant effects. The first effect is that heating decreases the contrast between the already existing absorbing regions and their background. These absorbing regions were produced by the Ly$\alpha$ coupling around the same sources. The sources continue to heat their surrounding IGM further and eventually convert these absorbing regions around them into heated regions. The brighter the X-ray sources are, the quicker this transition to generated heated regions around them happens. This implies that if the  $\fx$ value is large enough, one might observe this transition happening even during the {\em early} stage of the CD, compared to the comparatively fainter X-ray sources. This results either in a decrease in the magnitude of the negative bispectrum or the appearance of a positive bispectrum with increasing $\fx$. At the {\em early} stages, we also observe the decreasing bispectrum magnitude with increasing $\mh$ for a fixed $\fx$. This occurs since as
$\mh$ values increase the abundant low mass halos are no longer contributing to the state of the IGM. For the same reason the strength of the Ly$\alpha$ coupling also decreases (see the second panel in Figure \ref{fig:Tb_z}), resulting in a decrement in the magnitude of the negative bispectrum.

At the {\em intermediate} stages, we observe very drastic changes in the magnitude and sign of both squeezed-limit and equilateral bispectra in the $\fx - \mh$ parameter space, compared to the {\em early} stages. This is because, by the {\em intermediate} stage, the heating of the IGM has become a prominent physical process in almost all source models. The Ly$\alpha$ coupling, which has been the dominant IGM process so far, now competes with X-ray heating as the major contributor to the 21-cm fluctuations. The regions in the $\fx - \mh$ parameter space, where the bispectra for a particular triangle shape are negative, imply that this triangle shape probes the signal fluctuations dominated by the absorption regions. On the other hand, the positive bispectra imply that the signal fluctuations are dominated by the heated regions. 

The evolution of the bispectrum in the parameter space can be connected to the dominant IGM  physical processes in the following manner. For instance, at the {\em intermediate} stages, the squeezed-limit bispectrum for the lowest $\mh$ value shows a sign change, from negative to positive, as one changes the $\fx$ value slightly from its lowest value to a higher one. The magnitude of this positive bispectrum increases with increasing $\fx$ until the highest $\fx$ value is reached, where the bispectrum again becomes negative (i.e., there is a second sign change). This can be explained in the following way. The increase in $\fx$ increases the rate of heating by all the sources. In addition to this, for scenarios with low $\mh$ values, the low mass halos, which are numerous compared to the high mass halos, also contribute to the heating. These two effects together increase the overall heating rate. Hence, by the {\em intermediate} stages, the heated regions have been created in the Ly$\alpha$ coupled background in scenarios with higher $\fx$ values. Therefore, the bispectrum in these scenarios will be positive until one reaches the highest $\fx$ value. The heating rate for the highest $\fx$ value is high enough such that the heated regions start to overlap to form a single large connected heated cluster by the {\em intermediate} stage. Hence, the signal fluctuations at this stage can be thought of coming from the leftover absorption regions in a uniformly heated background (see the third panel from left in Figure \ref{fig:Tb_map} at $\fx = 1000$, and $\mh = 1\times10^9\, M_{\odot}$). The $\Delta_{\rm b}$, thus, will be negative, which results in a negative bispectrum. Further, for the highest $\mh$ values, the evolution trend of the squeezed-limit bispectrum with varying $\fx$ is quite different. The bispectrum magnitude first decreases with increasing $\fx$ until the largest $\fx$ value is reached, where it turns positive. This is because due to having the highest $\mh$ value in this scenario, the low-mass halos do not contribute at all to the heating process. Hence, the IGM heating is slow for $\fx = 0.1$, and the Ly$\alpha$ coupling is the sole major contributor to the signal fluctuations. Hence we get a large negative bispectrum in this scenario. As $\fx$ increases, the heated regions grow and gradually diminish the sizes of the absorption regions, hence their impact on the 21-cm fluctuations. This effectively explains the trend of the bispectrum discussed above.

During the {\em intermediate} stages, another interesting evolutionary feature of the bispectrum is observed with the variation in the $\mh$ for sources with the lowest values of $\fx$. This is the increasing magnitude of the negative bispectrum as one moves from the lowest values of $\mh$ towards its higher values. The physical interpretation of this feature is the following. The scenarios when the $\fx$ has a rather low value, even in them, the faint X-ray sources will start their contribution towards IGM heating by the {\em intermediate} stages of the CD. In such scenarios, in the cases with the lowest $\mh$, the sources will contribute significantly towards heating on top of the usual Ly$\alpha$ coupling. The heating will be faster for the lower $\mh$ values compared to the higher ones. Hence, the level of fluctuations introduced by the Ly$\alpha$ coupling to the 21-cm signal will decrease significantly by the {\em intermediate} stage. On the other hand, in scenarios with the highest $\mh$ values, the 21-cm signal fluctuations will be the highest due to the predominance of Ly$\alpha$ coupling. Hence, we observe the maximum negative bispectrum in these scenarios. The evolution of the bispectrum in the parameter space during the {\em late} stages of the CD can be similarly connected to the IGM physics.

The bispectrum of the equilateral triangle in the parameter space during the different CD stages are shown in the bottom panels of Figure \ref{fig:Mmin_fx_BS}. The evolution of the bispectrum for an equilateral triangle in the parameter space at a particular CD stage is more or less similar to the squeezed-limit bispectrum, with some differences. These differences are solely dependent on the shape of the $k$-triangle that probes a specific feature of the 21-cm field. For instance, at the {\em early} stage of the CD, the equilateral bispectrum for the sources with the lowest $\mh$ values and with increasing values of $\fx$ eventually show a sign change when one reaches the highest $\fx$ value. On the other hand, the squeezed-limit bispectrum does not show any such sign change for the same variations in the parameters. This is because the equilateral triangle can probe the fluctuations introduced by the heated region even at the very early stages of the CD. 

\subsection{Evolution of the bispectrum in the triangle parameter space}
\begin{figure}
  \centering
  \includegraphics[width=0.975\textwidth,angle=0]{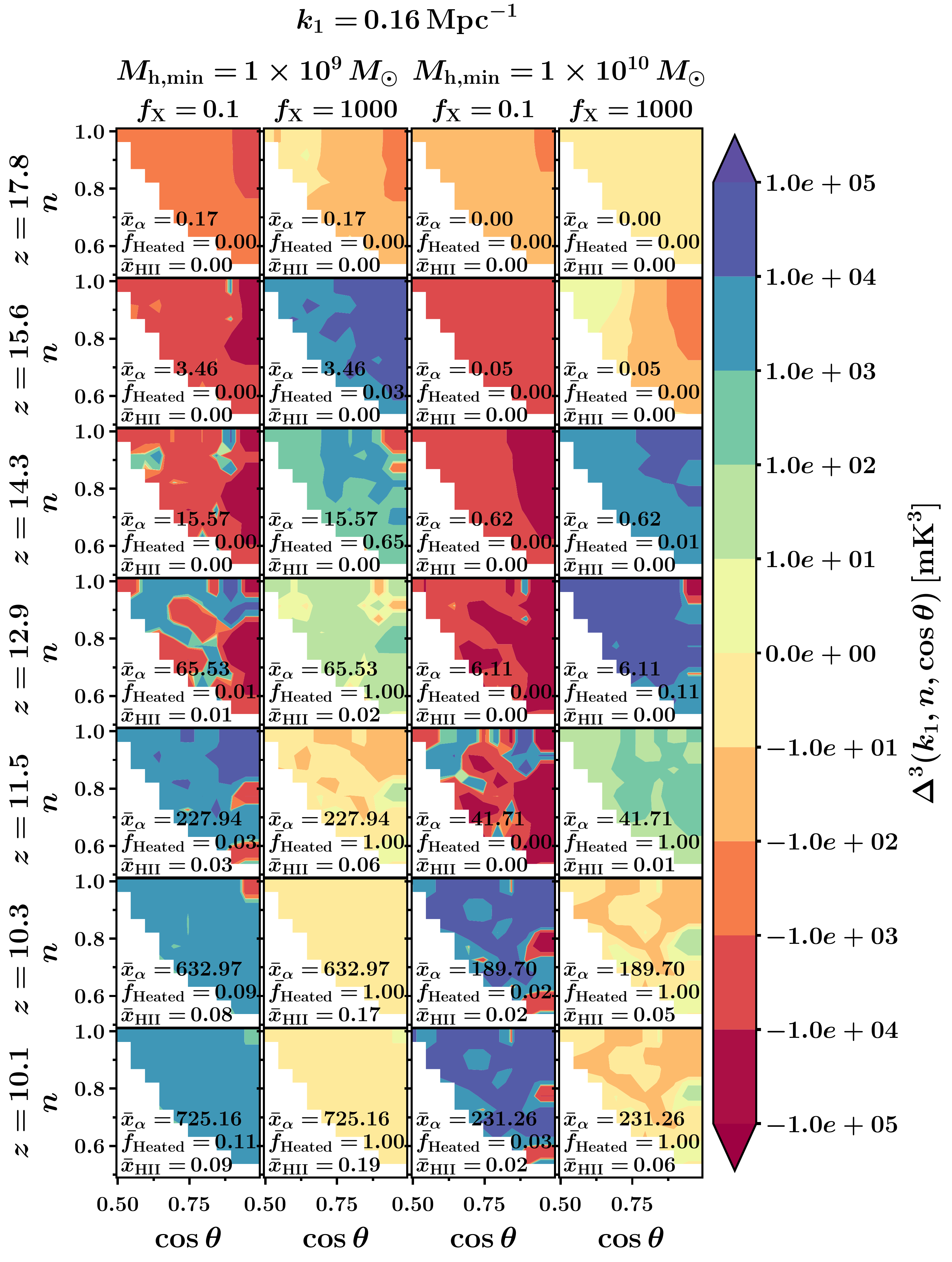}
  \caption{Shown are the 21-cm bispectra for all unique $k$-triangle configurations for $k_1 = 0.16 \mp$, for $\mh = 1\times10^9\, M_{\odot}$ (left two columns), and $\mh = 1\times10^{10}\, M_{\odot}$ (right two columns) at seven different stages of the CD and for two different $\fx$ values.}
  \label{fig:cost_n_BS}
\end{figure}

In this section, we discuss the bispectrum for all unique triangle shapes, focusing on two points: 1) to what extent the features of the bispectrum observed for squeezed-limit and equilateral $k$-triangles can be extended to their neighbouring triangle shapes in the $n$--$\cos{\theta}$ space; and 2) what more can be learned about the signal and the source models by considering the bispectrum for all unique triangles. Figure \ref{fig:cost_n_BS} shows the bispectra for all unique $k$-triangles with $k_1 = 0.16 \mp$. The first two columns in this figure show the bispectra for $\mh = 1\times10^{9}\, M_{\odot}$, at seven different stages of the CD and for two different $\fx$ values. The last two columns show the same for $\mh = 1\times10^{10}\, M_{\odot}$. During the very early stages of the CD, for any value of $\mh$ and $\fx$, the bispectra are negative in the entire unique $n$--$\cos{\theta}$ space (see the top panels in Figure \ref{fig:cost_n_BS}). This is because, for any CD scenario, the heating of the IGM is negligible at this stage, and the Ly$\alpha$ coupling dictates 21-cm fluctuations. Thus any of the unique $k$-triangles probes the fluctuations introduced by the absorbing regions.

In Section \ref{sec:sq_eq_bs} we have reported that, for the scenario with $\mh = 1\times10^{9}\, M_{\odot}$ and $\fx = 0.1$, the redshift evolution of the squeezed-limit bispectrum shows a total of three sign changes (see solid black line in the top-left panel of Figure \ref{fig:BS_z_eq_sq}). In the unique $n$--$\cos{\theta}$ space, we observe the same features for a set of $k$-triangles that are in the vicinity of the squeezed $k$-triangle (see the left-most column of Figure \ref{fig:cost_n_BS}). Furthermore, for the same $\mh$ and $\fx$ values, we have also reported the single sign change in the equilateral bispectrum (see solid black line in the bottom-left panel of Figure \ref{fig:BS_z_eq_sq}). Most of the $k$-triangles in the unique $n$--$\cos{\theta}$ space show this feature, except the ones in the vicinity of the squeezed-limit triangles. Another interesting observation based on this analysis is that the redshifts where these sign changes of the  bispectrum occur depends on the triangle shapes. For a few triangles that do not belong to the L-isosceles group, these sign changes happen at an earlier redshift compared to the squeezed-limit and equilateral triangles. This is because the bispectrum estimation for these triangles involves signal fluctuations coming from multiple length scales; thus, they may be able to pick up the impact of heating even earlier than the squeezed-limit bispectrum. Further, as we increase the value of $\mh$ to $10^{10}\, M_{\odot}$ and keep $\fx$ fixed at $0.1$, we observe that the same sign change features appear in the bispectrum. The only difference is that in this case, because of the lack of source hosting halos early on, these features appear at later redshifts during the CD. 

In Section \ref{sec:sq_eq_bs} we have also reported that, for the scenarios with $\fx = 1000$ and for any value of $\mh$, the evolution of the squeezed-limit bispectrum with redshift show a total of four sign changes (see the top-right panel of Figure \ref{fig:BS_z_eq_sq}). We observe the same features in the 21-cm bispectra for a set of $k$-triangles that are in the vicinity of the squeezed $k$-triangle in the unique $n$--$\cos{\theta}$ space (see panels in second and fourth columns from the left of Figure \ref{fig:cost_n_BS}). Further, the equilateral bispectrum shows a double sign change (see the bottom-right panel of Figure \ref{fig:BS_z_eq_sq}). Most of the $k$-triangles in the unique $n$--$\cos{\theta}$ space show the same 21-cm bispectra feature, except the ones in the vicinity of squeezed-limit triangles. When we compare all of these features obtained for scenarios with two extreme values of $\mh$, the features observed for the largest $\mh$ value appear to be delayed compared to the same for the smallest $\mh$ value.

\section{Summary and discussions}
\label{sec:summary}
This article is a follow-up to our previous work \citep{kamran21}. In \citep{kamran21}, we have shown how the 21-cm signal bispectrum probes the impact of all possible astrophysical processes on the signal fluctuations by capturing the intrinsic non-Gaussianity in the signal during Cosmic Dawn (CD). The Ly$\alpha$ coupling and X-ray heating are the two dominant astrophysical processes during CD. The third process, photo-ionization, becomes important during the late stages of the CD. In \citep{kamran21}, we have considered that the radiating sources reside inside the halos whose masses are above a certain minimum threshold ($\mh$). To probe the signature of the aforementioned physical processes in the IGM induced by these radiating sources, we considered several CD scenarios. In a few of these scenarios, there were no X-ray emission from the sources (i.e., $\fx = 0$). On the other hand, the sources corresponding to the rest of the scenarios emit ample amounts of X-ray photons (i.e. for which the $\fx$ values are high). However, in \citep{kamran21}, we have not studied how the different X-ray sources, from fainter to brighter, impact the 21-cm signal and its bispectrum through the possible astrophysical processes. We have also not studied the effect of different $\mh$ values on the 21-cm signal and its bispectrum. These CD scenarios impact the IGM with the different Ly$\alpha$ coupling, X-ray heating, and photo-ionization processes. Besides these, we studied only the squeezed-limit bispectrum in the earlier work.

The current work focuses on addressing the following key questions: 1) to what extent and how varying source parameter values impact the CD signal bispectrum; 2) to what extent the results for squeezed and equilateral bispectra can be extended to bispectrum obtained for triangles of other unique shapes; and 3) what additional information may be extracted using the bispectrum of various unique triangle shapes. In this work, therefore, we have considered several simulated CD scenarios corresponding to all possible combinations of the values of the source parameters, $\fx$ and $\mh$, within certain ranges (see Table \ref{table1}). We focus on how these different CD scenarios that affect the 21-cm signal via the different astrophysical processes will impact the large-scale ($k_1 = 0.16 \mp$) 21-cm bispectrum. The findings of our analysis can be summarized as follows:

\begin{itemize}
    \item The different astrophysical source parameters during CD, such as $\mh$ and $\fx$, directly control the physical processes going on in the IGM. These processes determine the nature of fluctuations in the 21-cm signal emerging from the IGM. One can relate the fluctuations in the signal to the IGM processes and thus to the source parameters to some extent via the signal power spectrum. However, the signal bispectrum provides a far better and more robust connection between the signal fluctuations and the source parameters. This is because compared to the power spectrum, which is always positive by definition, the bispectrum can have both positive and negative values. Furthermore, the shape of the bispectrum is also more sensitive to the variation in these source parameters. This is due to the simple fact that the bispectrum is capable of quantifying the intrinsic non-Gaussianity in the signal to which the power spectrum is not sensitive. Therefore both the sign and shape of the bispectrum work as a more sensitive smoking gun for the ongoing dominant physical processes in the IGM induced by the specific source properties (or parameters). 
    
    \item The nature and the level of non-Gaussianity in the 21-cm signal emerging from the IGM depend on the nature of the sources (defined by their parameter values). The nature of the sources determines which physical process in the IGM is dominant at what cosmic time. For example, we observed in our analysis that, if we consider two different CD scenarios, both having the same values of $\mh$ but different values of $\fx$, then the one with lower $\fx$ (i.e. having fainter X-ray photon producing sources) will be able to heat up the entire IGM at a far later stage compared to the scenario which has higher $\fx$ values. These two scenarios can be conclusively distinguished by the shape and sign (and their variation with cosmic time) of the bispectra estimated using small $k$ (i.e. large length scale) squeezed limit triangles or triangle shapes that are close to it in the $n$--$\cos{\theta}$ space.
    
    \item Next, let us consider two CD scenarios with the same $\fx$ values but different $\mh$ values. In this case, sources of equal masses in both scenarios produce an equal amount of X-ray photons. However, the scenario where $\mh$ is lower will have a larger number of low mass sources. This results in a faster IGM heating in the scenario with lower $\mh$ values. Therefore, the features in the bispectrum connected to the dominance of X-ray heating will appear at an earlier redshift in the scenario with lower $\mh$ compared to the other one. Thus the evolution of the bispectra with cosmic time will appear to be shifted with respect to each other in these two CD scenarios while keeping the nature and its prominent features almost the same. Similar to the discussion in the previous point, here also we find that bispectrum for small $k$ (i.e. large length scale) squeezed limit triangles and the triangle shapes in its vicinity are able to optimally probe the impact of X-ray heating even in this case.
    
    \item The analysis presented in this paper, both in terms of the simulations of the signal and our physical interpretation of the bispectra of the simulated signal, is based on the assumption that the first dominant IGM physical process when the first sources of lights were formed was Ly$\alpha$ coupling. These sources started to impact the IGM via X-ray heating at a later time, depending on their corresponding $\fx$ and $\mh$ values. Thus after analyzing the signal bispectra for a large number of $\fx$ and $\mh$ values, we arrive at the following generic conclusion: The sign of the 21-cm bispectrum for a particular triangle configuration can tell us the relative contrast of the fluctuations in the 21-cm signal with respect to its background. For example, a negative bispectrum can arise under two conditions in the IGM. First, when the fluctuations in the signal are dominated by the distribution of cold absorbing regions in a relatively warm or hot background. Second, when the signal fluctuations are determined by the distribution of heated regions in a relatively cold or less warm background. Therefore, just by looking at the sign of the bispectrum, it would be difficult to say which among the two conditions mentioned above is resulting in the negative bispectrum. In this paper, we demonstrate using the suite of signal simulations at our disposal, that it is important to study the sequence of sign changes along with the variations in the shape and magnitude of the bispectrum throughout the CD history to arrive at a robust conclusion about the dominant IGM process at different cosmic times.  
\end{itemize}

We do not consider CD/EoR models where photo-ionization already becomes important during the CD, since those seem unlikely in view of our current knowledge. Instead, we only considered the effects of the Ly$\alpha$ coupling and X-ray heating processes, but we do include the impact of photo-ionization during the late stage of the CD in order to see how the 21-cm signal and its bispectrum gradually transit to the EoR. Furthermore, we did not consider here how any residual foregrounds may affect the 21-cm bispectrum. It is vital to use the techniques of the optimal foreground subtraction for a reliable estimation of the signal bispectra from the observed data \cite{watkinson20,hothi20}. Improper foreground subtraction may lead to a wrong interpretation of the signal bispectra. We have also not made the detectability predictions of the 21-cm bispectra for any currently operating or upcoming telescopes by considering the presence of thermal noise \cite{yoshiura15} and other systematic uncertainties. We plan to address these issues in our follow-up work.

\section{Acknowledgements}
MK is supported by the foundation Carl Tryggers stiftelse för vetenskaplig forskning, under grant agreement 21:1376 awarded to docent Martin Sahlén. 
SM acknowledges financial support through the project titled ``Observing the Cosmic Dawn in Multicolour using Next Generation Telescopes'' funded by the Science and Engineering Research Board (SERB), Department of Science and Technology, Government of India through the Core Research Grant No. CRG/2021/004025.
RM is supported by the Israel Academy of Sciences and Humanities \& Council for Higher Education Excellence Fellowship Program for International Postdoctoral Researchers. RG acknowledges support by the Israel Science Foundation grant no. 255/18. GM is supported by Swedish Research Council grant
2020-04691.
This work is also supported by the Science and Technology Facilities Council [grant numbers ST/I000976/1, ST/P000525/1 and ST/T000473/1] and the Southeast Physics Network (SEPNet). The authors gratefully acknowledge the Gauss Centre for Supercomputing e.V. (\url{www.gauss-centre.eu}) for funding this project by providing computing time through the John von Neumann Institute for Computing (NIC) on the GCS Supercomputer JUWELS at Jülich Supercomputing Centre (JSC).We acknowledge that the results of this research have been achieved in part using the DECI resource Beskow based in Sweden at PDC with support from the PRACE aisbl.

\appendix

\section{Evolution of the CD 21-cm bispectra for intermediate and small length scales}
\label{sec:BS_int_small_k1}
\begin{figure*}
  \includegraphics[width=1\textwidth,angle=0]{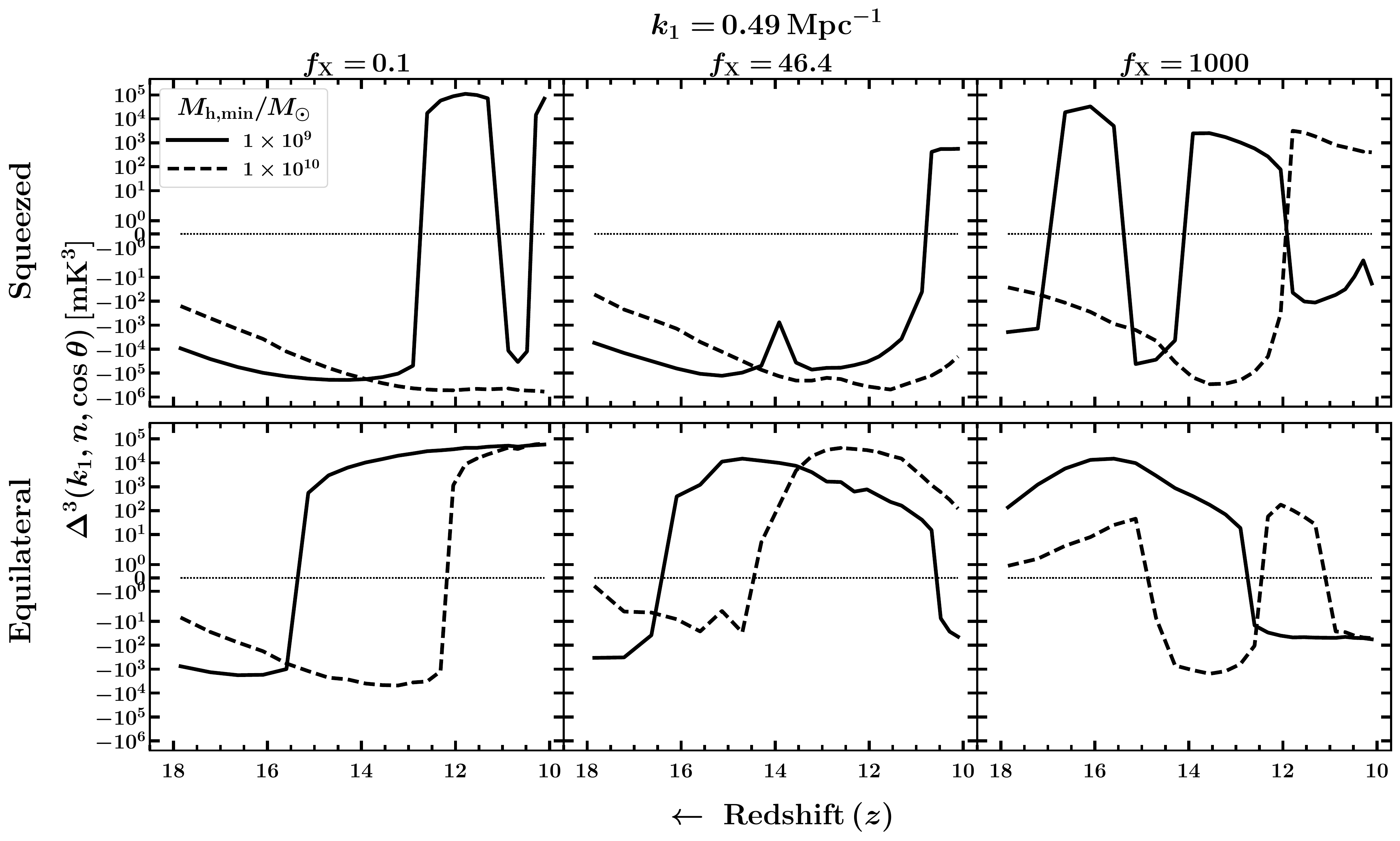}
  \caption{Shown is the evolution of the bispectra at $k_1 = 0.49 \mp$ in (top) for squeezed limit; and (bottom) for equilateral $k$-triangles. For each panel the bispectra are presented at a fixed $\fx$, but with two extreme values of $\mh$.}
  \label{fig:BS_z_eq_sq_k1_0.5}
\end{figure*}

\begin{figure*}
  \includegraphics[width=1\textwidth,angle=0]{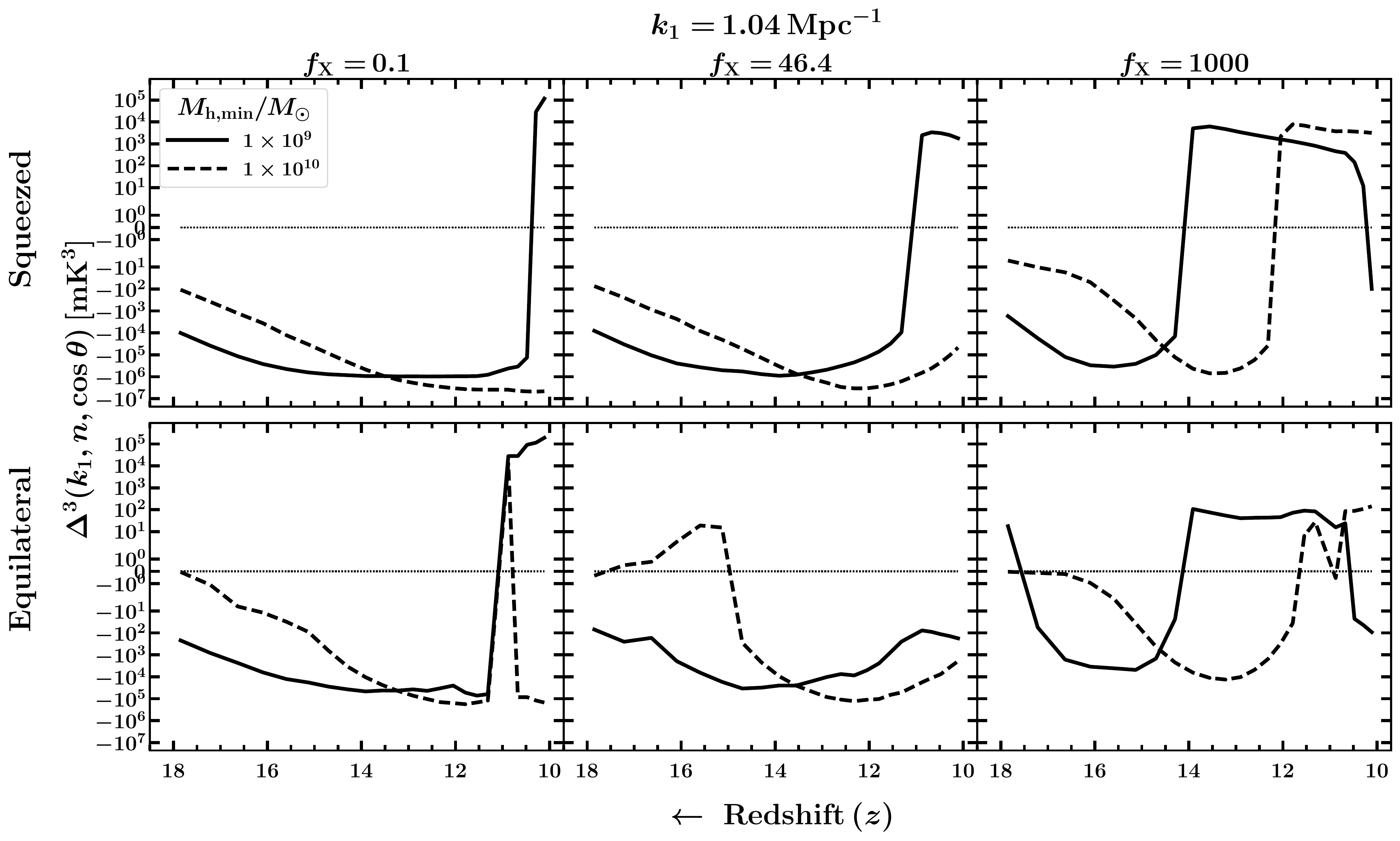}
  \caption{Same as in Fig.~\ref{fig:BS_z_eq_sq_k1_0.5}, but at
  $k_1 = 1.04 \mp$.}
  \label{fig:BS_z_eq_sq_k1_1}
\end{figure*}

As mentioned earlier, our main focus in this paper is to study the redshift evolution of the large scale 21-cm bispectra for various source models. Thus most of our discussion is concentrated on bispectra for $k_1 = 0.16 \mp$. However, for the sake of completeness, we also show the bispectrum evolution for intermediate ($k_1 = 0.49 \mp$) and small ($k_1 = 1.04 \mp$) length scales in Figures \ref{fig:BS_z_eq_sq_k1_0.5} and  \ref{fig:BS_z_eq_sq_k1_1}. The evolution of the intermediate length scale bispectrum is somewhat similar to the large scale bispectrum as observed in Figure \ref{fig:BS_z_eq_sq} and can be interpreted via a similar set of arguments as discussed in Section \ref{sec:sq_eq_bs}. However, the evolution of the small length scale bispectrum differs significantly from that of the large and intermediate length scale bispectra and requires a more through analysis which we plan to take up in a future project.

\bibliographystyle{JHEP}
\bibliography{ref}

\providecommand{\href}[2]{#2}\begingroup\raggedright\begin{thebibliography}{10}

\bibitem{barkana01}
R.~{Barkana} and A.~{Loeb}, \emph{{In the beginning: the first sources of light
  and the reionization of the universe}}, {\emph{Physics Reports} {\bfseries
  349} (2001) 125}
  [\href{https://arxiv.org/abs/arXiv:astro-ph/0010468}{{\ttfamily
  arXiv:astro-ph/0010468}}].

\bibitem{furlanetto06}
S.R.~{Furlanetto}, S.P.~{Oh} and F.H.~{Briggs}, \emph{{Cosmology at low
  frequencies: The 21 cm transition and the high-redshift Universe}},
  {\emph{Physics Reports} {\bfseries 433} (2006) 181}
  [\href{https://arxiv.org/abs/arXiv:astro-ph/0608032}{{\ttfamily
  arXiv:astro-ph/0608032}}].

\bibitem{pritchard12}
J.R.~{Pritchard} and A.~{Loeb}, \emph{{21 cm cosmology in the 21st century}},
  {\emph{Reports on Progress in Physics} {\bfseries 75} (2012) 086901}
  [\href{https://arxiv.org/abs/1109.6012}{{\ttfamily 1109.6012}}].

\bibitem{paciga13}
G.~{Paciga}, J.G.~{Albert}, K.~{Bandura}, T.-C.~{Chang}, Y.~{Gupta},
  C.~{Hirata} et~al., \emph{{A simulation-calibrated limit on the H I power
  spectrum from the GMRT Epoch of Reionization experiment}},
  \href{https://doi.org/10.1093/mnras/stt753}{\emph{\mnras} {\bfseries 433}
  (2013) 639} [\href{https://arxiv.org/abs/1301.5906}{{\ttfamily 1301.5906}}].

\bibitem{mertens20}
F.G.~Mertens, M.~Mevius, L.V.E.~Koopmans, A.R.~Offringa, G.~Mellema, S.~Zaroubi
  et~al., \emph{{Improved upper limits on the 21 cm signal power spectrum of
  neutral hydrogen at z ≈ 9.1 from LOFAR}},
  \href{https://doi.org/10.1093/mnras/staa327}{\emph{Monthly Notices of the
  Royal Astronomical Society} {\bfseries 493} (2020) 1662}
  [\href{https://arxiv.org/abs/https://academic.oup.com/mnras/article-pdf/493/2/1662/32666766/staa327.pdf}{{\ttfamily
  https://academic.oup.com/mnras/article-pdf/493/2/1662/32666766/staa327.pdf}}].

\bibitem{barry19}
N.~{Barry}, M.~{Wilensky}, C.M.~{Trott}, B.~{Pindor}, A.P.~{Beardsley},
  B.J.~{Hazelton} et~al., \emph{{Improving the Epoch of Reionization Power
  Spectrum Results from Murchison Widefield Array Season 1 Observations}},
  \href{https://doi.org/10.3847/1538-4357/ab40a8}{\emph{\apj} {\bfseries 884}
  (2019) 1} [\href{https://arxiv.org/abs/1909.00561}{{\ttfamily 1909.00561}}].

\bibitem{kolopanis19}
M.~Kolopanis, D.~Jacobs, C.~Cheng, A.~Parsons, S.~Kohn, J.~Pober et~al.,
  \emph{A simplified, lossless reanalysis of paper-64},
  \href{https://doi.org/10.3847/1538-4357/ab3e3a}{\emph{The Astrophysical
  Journal} {\bfseries 883} (2019) 133}.

\bibitem{deboer17}
D.R.~DeBoer, A.R.~Parsons, J.E.~Aguirre, P.~Alexander, Z.S.~Ali, A.P.~Beardsley
  et~al., \emph{Hydrogen epoch of reionization array ({HERA})},
  \href{https://doi.org/10.1088/1538-3873/129/974/045001}{\emph{Publications of
  the Astronomical Society of the Pacific} {\bfseries 129} (2017) 045001}.

\bibitem{HERA21}
{The HERA Collaboration}, Z.~{Abdurashidova}, J.E.~{Aguirre}, P.~{Alexander},
  Z.S.~{Ali}, Y.~{Balfour} et~al., \emph{{First Results from HERA Phase I:
  Upper Limits on the Epoch of Reionization 21 cm Power Spectrum}},
  {\emph{arXiv e-prints} (2021) arXiv:2108.02263}
  [\href{https://arxiv.org/abs/2108.02263}{{\ttfamily 2108.02263}}].

\bibitem{2016ApJ...818..139T}
C.M.~{Trott}, B.~{Pindor}, P.~{Procopio}, R.B.~{Wayth}, D.A.~{Mitchell},
  B.~{McKinley} et~al., \emph{{CHIPS: The Cosmological H I Power Spectrum
  Estimator}}, \href{https://doi.org/10.3847/0004-637X/818/2/139}{\emph{\apj}
  {\bfseries 818} (2016) 139}
  [\href{https://arxiv.org/abs/1601.02073}{{\ttfamily 1601.02073}}].

\bibitem{2021MNRAS.500.2264H}
I.~{Hothi}, E.~{Chapman}, J.R.~{Pritchard}, F.G.~{Mertens}, L.V.E.~{Koopmans},
  B.~{Ciardi} et~al., \emph{{Comparing foreground removal techniques for
  recovery of the LOFAR-EoR 21 cm power spectrum}},
  \href{https://doi.org/10.1093/mnras/staa3446}{\emph{\mnras} {\bfseries 500}
  (2021) 2264} [\href{https://arxiv.org/abs/2011.01284}{{\ttfamily
  2011.01284}}].

\bibitem{2022MNRAS.509.3693M}
M.~{Mevius}, F.~{Mertens}, L.V.E.~{Koopmans}, A.R.~{Offringa}, S.~{Yatawatta},
  M.A.~{Brentjens} et~al., \emph{{A numerical study of 21-cm signal suppression
  and noise increase in direction-dependent calibration of LOFAR data}},
  \href{https://doi.org/10.1093/mnras/stab3233}{\emph{\mnras} {\bfseries 509}
  (2022) 3693} [\href{https://arxiv.org/abs/2111.02537}{{\ttfamily
  2111.02537}}].

\bibitem{2022arXiv220302345G}
H.~{Gan}, L.V.~{E Koopmans}, F.G.~{Mertens}, M.~{Mevius}, A.R.~{Offringa},
  B.~{Ciardi} et~al., \emph{{Statistical analysis of the causes of excess
  variance in the 21 cm signal power spectra obtained with the Low-Frequency
  Array}}, {\emph{arXiv e-prints} (2022) arXiv:2203.02345}
  [\href{https://arxiv.org/abs/2203.02345}{{\ttfamily 2203.02345}}].

\bibitem{iliev08}
I.T.~{Iliev}, G.~{Mellema}, U.-L.~{Pen}, J.R.~{Bond} and P.R.~{Shapiro},
  \emph{{Current models of the observable consequences of cosmic reionization
  and their detectability}}, {\emph{\mnras} {\bfseries 384} (2008) 863}
  [\href{https://arxiv.org/abs/arXiv:astro-ph/0702099}{{\ttfamily
  arXiv:astro-ph/0702099}}].

\bibitem{patil14}
A.H.~{Patil}, S.~{Zaroubi}, E.~{Chapman}, V.~{Jeli{\'c}}, G.~{Harker},
  F.B.~{Abdalla} et~al., \emph{{Constraining the epoch of reionization with the
  variance statistic: simulations of the LOFAR case}},
  \href{https://doi.org/10.1093/mnras/stu1178}{\emph{\mnras} {\bfseries 443}
  (2014) 1113} [\href{https://arxiv.org/abs/1401.4172}{{\ttfamily 1401.4172}}].

\bibitem{watkinson15}
C.A.~{Watkinson} and J.R.~{Pritchard}, \emph{{The impact of spin-temperature
  fluctuations on the 21-cm moments}},
  \href{https://doi.org/10.1093/mnras/stv2010}{\emph{\mnras} {\bfseries 454}
  (2015) 1416} [\href{https://arxiv.org/abs/1505.07108}{{\ttfamily
  1505.07108}}].

\bibitem{ali08}
S.S.~{Ali}, S.~{Bharadwaj} and J.N.~{Chengalur}, \emph{{Foregrounds for
  redshifted 21-cm studies of reionization: Giant Meter Wave Radio Telescope
  153-MHz observations}}, {\emph{\mnras} {\bfseries 385} (2008) 2166}
  [\href{https://arxiv.org/abs/0801.2424}{{\ttfamily 0801.2424}}].

\bibitem{ghosh12}
A.~{Ghosh}, J.~{Prasad}, S.~{Bharadwaj}, S.S.~{Ali} and J.N.~{Chengalur},
  \emph{{Characterizing foreground for redshifted 21 cm radiation: 150 MHz
  Giant Metrewave Radio Telescope observations}},
  \href{https://doi.org/10.1111/j.1365-2966.2012.21889.x}{\emph{\mnras}
  {\bfseries 426} (2012) 3295}
  [\href{https://arxiv.org/abs/1208.1617}{{\ttfamily 1208.1617}}].

\bibitem{li19}
W.~{Li}, J.C.~{Pober}, N.~{Barry}, B.J.~{Hazelton}, M.F.~{Morales},
  C.M.~{Trott} et~al., \emph{{First Season MWA Phase II Epoch of Reionization
  Power Spectrum Results at Redshift 7}},
  \href{https://doi.org/10.3847/1538-4357/ab55e4}{\emph{\apj} {\bfseries 887}
  (2019) 141} [\href{https://arxiv.org/abs/1911.10216}{{\ttfamily
  1911.10216}}].

\bibitem{trott20}
C.M.~Trott, C.H.~Jordan, S.~Midgley, N.~Barry, B.~Greig, B.~Pindor et~al.,
  \emph{{Deep multiredshift limits on Epoch of Reionization 21 cm power
  spectra from four seasons of Murchison Widefield Array observations}},
  \href{https://doi.org/10.1093/mnras/staa414}{\emph{Monthly Notices of the
  Royal Astronomical Society} {\bfseries 493} (2020) 4711}
  [\href{https://arxiv.org/abs/https://academic.oup.com/mnras/article-pdf/493/4/4711/32927265/staa414.pdf}{{\ttfamily
  https://academic.oup.com/mnras/article-pdf/493/4/4711/32927265/staa414.pdf}}].

\bibitem{mellema15}
G.~{Mellema}, L.~{Koopmans}, H.~{Shukla}, K.K.~{Datta}, A.~{Mesinger} and
  S.~{Majumdar}, \emph{{HI tomographic imaging of the Cosmic Dawn and Epoch of
  Reionization with SKA}}, {\emph{Advancing Astrophysics with the Square
  Kilometre Array (AASKA14)} (2015) 10}
  [\href{https://arxiv.org/abs/1501.04203}{{\ttfamily 1501.04203}}].

\bibitem{koopmans15}
L.~{Koopmans}, J.~{Pritchard}, G.~{Mellema}, J.~{Aguirre}, K.~{Ahn},
  R.~{Barkana} et~al., \emph{{The Cosmic Dawn and Epoch of Reionisation with
  SKA}}, {\emph{Advancing Astrophysics with the Square Kilometre Array
  (AASKA14)} (2015) 1} [\href{https://arxiv.org/abs/1505.07568}{{\ttfamily
  1505.07568}}].

\bibitem{ghara16}
R.~{Ghara}, T.R.~{Choudhury} and K.K.~{Datta}, \emph{{21-cm signature of the
  first sources in the Universe: prospects of detection with SKA}},
  \href{https://doi.org/10.1093/mnras/stw953}{\emph{\mnras} {\bfseries 460}
  (2016) 827} [\href{https://arxiv.org/abs/1511.07448}{{\ttfamily
  1511.07448}}].

\bibitem{bharadwaj04}
S.~{Bharadwaj} and S.S.~{Ali}, \emph{{The cosmic microwave background radiation
  fluctuations from HI perturbations prior to reionization}}, {\emph{\mnras}
  {\bfseries 352} (2004) 142}
  [\href{https://arxiv.org/abs/arXiv:astro-ph/0401206}{{\ttfamily
  arXiv:astro-ph/0401206}}].

\bibitem{bharadwaj05}
S.~{Bharadwaj} and S.S.~{Ali}, \emph{{On using visibility correlations to probe
  the HI distribution from the dark ages to the present epoch - I. Formalism
  and the expected signal}}, {\emph{\mnras} {\bfseries 356} (2005) 1519}
  [\href{https://arxiv.org/abs/arXiv:astro-ph/0406676}{{\ttfamily
  arXiv:astro-ph/0406676}}].

\bibitem{barkana05}
R.~{Barkana} and A.~{Loeb}, \emph{{A Method for Separating the Physics from the
  Astrophysics of High-Redshift 21 Centimeter Fluctuations}}, {\emph{\apjl}
  {\bfseries 624} (2005) L65}
  [\href{https://arxiv.org/abs/arXiv:astro-ph/0409572}{{\ttfamily
  arXiv:astro-ph/0409572}}].

\bibitem{datta07a}
K.K.~{Datta}, T.R.~{Choudhury} and S.~{Bharadwaj}, \emph{{The multifrequency
  angular power spectrum of the epoch of reionization 21-cm signal}},
  {\emph{\mnras} {\bfseries 378} (2007) 119}
  [\href{https://arxiv.org/abs/arXiv:astro-ph/0605546}{{\ttfamily
  arXiv:astro-ph/0605546}}].

\bibitem{datta14}
K.K.~{Datta}, H.~{Jensen}, S.~{Majumdar}, G.~{Mellema}, I.T.~{Iliev}, Y.~{Mao}
  et~al., \emph{{Light cone effect on the reionization 21-cm signal - II.
  Evolution, anisotropies and observational implications}},
  \href{https://doi.org/10.1093/mnras/stu927}{\emph{\mnras} {\bfseries 442}
  (2014) 1491} [\href{https://arxiv.org/abs/1402.0508}{{\ttfamily 1402.0508}}].

\bibitem{mesinger07}
A.~{Mesinger} and S.~{Furlanetto}, \emph{{Efficient Simulations of Early
  Structure Formation and Reionization}}, {\emph{\apj} {\bfseries 669} (2007)
  663} [\href{https://arxiv.org/abs/0704.0946}{{\ttfamily 0704.0946}}].

\bibitem{lidz08}
A.~{Lidz}, O.~{Zahn}, M.~{McQuinn}, M.~{Zaldarriaga} and L.~{Hernquist},
  \emph{{Detecting the Rise and Fall of 21 cm Fluctuations with the Murchison
  Widefield Array}}, {\emph{\apj} {\bfseries 680} (2008) 962}
  [\href{https://arxiv.org/abs/0711.4373}{{\ttfamily 0711.4373}}].

\bibitem{choudhury09b}
T.R.~{Choudhury}, M.G.~{Haehnelt} and J.~{Regan}, \emph{{Inside-out or
  outside-in: the topology of reionization in the photon-starved regime
  suggested by Ly{$\alpha$} forest data}}, {\emph{\mnras} {\bfseries 394}
  (2009) 960} [\href{https://arxiv.org/abs/0806.1524}{{\ttfamily 0806.1524}}].

\bibitem{mao12}
Y.~{Mao}, P.R.~{Shapiro}, G.~{Mellema}, I.T.~{Iliev}, J.~{Koda} and K.~{Ahn},
  \emph{{Redshift-space distortion of the 21-cm background from the epoch of
  reionization - I. Methodology re-examined}}, {\emph{\mnras} {\bfseries 422}
  (2012) 926} [\href{https://arxiv.org/abs/1104.2094}{{\ttfamily 1104.2094}}].

\bibitem{jensen13}
H.~{Jensen}, K.K.~{Datta}, G.~{Mellema}, E.~{Chapman}, F.B.~{Abdalla},
  I.T.~{Iliev} et~al., \emph{{Probing reionization with LOFAR using 21-cm
  redshift space distortions}},
  \href{https://doi.org/10.1093/mnras/stt1341}{\emph{\mnras} {\bfseries 435}
  (2013) 460} [\href{https://arxiv.org/abs/1303.5627}{{\ttfamily 1303.5627}}].

\bibitem{majumdar13}
S.~{Majumdar}, S.~{Bharadwaj} and T.R.~{Choudhury}, \emph{{The effect of
  peculiar velocities on the epoch of reionization 21-cm signal}},
  \href{https://doi.org/10.1093/mnras/stt1144}{\emph{\mnras} {\bfseries 434}
  (2013) 1978} [\href{https://arxiv.org/abs/1209.4762}{{\ttfamily 1209.4762}}].

\bibitem{majumdar14}
S.~{Majumdar}, G.~{Mellema}, K.K.~{Datta}, H.~{Jensen}, T.R.~{Choudhury},
  S.~{Bharadwaj} et~al., \emph{{On the use of seminumerical simulations in
  predicting the 21-cm signal from the epoch of reionization}},
  \href{https://doi.org/10.1093/mnras/stu1342}{\emph{\mnras} {\bfseries 443}
  (2014) 2843} [\href{https://arxiv.org/abs/1403.0941}{{\ttfamily 1403.0941}}].

\bibitem{mondal15}
R.~{Mondal}, S.~{Bharadwaj}, S.~{Majumdar}, A.~{Bera} and A.~{Acharyya},
  \emph{{The effect of non-Gaussianity on error predictions for the Epoch of
  Reionization (EoR) 21-cm power spectrum}},
  \href{https://doi.org/10.1093/mnrasl/slv015}{\emph{\mnras} {\bfseries 449}
  (2015) L41} [\href{https://arxiv.org/abs/1409.4420}{{\ttfamily 1409.4420}}].

\bibitem{mondal16}
R.~{Mondal}, S.~{Bharadwaj} and S.~{Majumdar}, \emph{{Statistics of the epoch
  of reionization 21-cm signal - I. Power spectrum error-covariance}},
  \href{https://doi.org/10.1093/mnras/stv2772}{\emph{\mnras} {\bfseries 456}
  (2016) 1936} [\href{https://arxiv.org/abs/1508.00896}{{\ttfamily
  1508.00896}}].

\bibitem{majumdar16}
S.~{Majumdar}, H.~{Jensen}, G.~{Mellema}, E.~{Chapman}, F.B.~{Abdalla},
  K.-Y.~{Lee} et~al., \emph{{Effects of the sources of reionization on 21-cm
  redshift-space distortions}},
  \href{https://doi.org/10.1093/mnras/stv2812}{\emph{\mnras} {\bfseries 456}
  (2016) 2080} [\href{https://arxiv.org/abs/1509.07518}{{\ttfamily
  1509.07518}}].

\bibitem{mondal17}
R.~{Mondal}, S.~{Bharadwaj} and S.~{Majumdar}, \emph{{Statistics of the epoch
  of reionization (EoR) 21-cm signal - II. The evolution of the power-spectrum
  error-covariance}},
  \href{https://doi.org/10.1093/mnras/stw2599}{\emph{\mnras} {\bfseries 464}
  (2017) 2992} [\href{https://arxiv.org/abs/1606.03874}{{\ttfamily
  1606.03874}}].

\bibitem{bharadwaj05a}
S.~{Bharadwaj} and S.K.~{Pandey}, \emph{{Probing non-Gaussian features in the
  HI distribution at the epoch of re-ionization}}, {\emph{\mnras} {\bfseries
  358} (2005) 968}
  [\href{https://arxiv.org/abs/arXiv:astro-ph/0410581}{{\ttfamily
  arXiv:astro-ph/0410581}}].

\bibitem{mellema06}
G.~{Mellema}, I.T.~{Iliev}, U.-L.~{Pen} and P.R.~{Shapiro}, \emph{{Simulating
  cosmic reionization at large scales - II. The 21-cm emission features and
  statistical signals}}, {\emph{\mnras} {\bfseries 372} (2006) 679}
  [\href{https://arxiv.org/abs/arXiv:astro-ph/0603518}{{\ttfamily
  arXiv:astro-ph/0603518}}].

\bibitem{kamran20}
M.~{Kamran}, R.~{Ghara}, S.~{Majumdar}, R.~{Mondal}, G.~{Mellema},
  S.~{Bharadwaj} et~al., \emph{{Redshifted 21-cm bispectrum - II. Impact of the
  spin temperature fluctuations and redshift space distortions on the signal
  from the Cosmic Dawn}},
  \href{https://doi.org/10.1093/mnras/stab216}{\emph{\mnras} {\bfseries 502}
  (2021) 3800} [\href{https://arxiv.org/abs/2012.11616}{{\ttfamily
  2012.11616}}].

\bibitem{mondal21}
R.~{Mondal}, G.~{Mellema}, A.K.~{Shaw}, M.~{Kamran} and S.~{Majumdar},
  \emph{{The Epoch of Reionization 21-cm bispectrum: the impact of light-cone
  effects and detectability}},
  \href{https://doi.org/10.1093/mnras/stab2900}{\emph{\mnras} {\bfseries 508}
  (2021) 3848} [\href{https://arxiv.org/abs/2107.02668}{{\ttfamily
  2107.02668}}].

\bibitem{kamran21}
M.~{Kamran}, S.~{Majumdar}, R.~{Ghara}, G.~{Mellema}, S.~{Bharadwaj},
  J.R.~{Pritchard} et~al., \emph{{Probing IGM Physics during Cosmic Dawn using
  the Redshifted 21-cm Bispectrum}}, {\emph{arXiv e-prints} (2021)
  arXiv:2108.08201} [\href{https://arxiv.org/abs/2108.08201}{{\ttfamily
  2108.08201}}].

\bibitem{shaw19}
A.K.~{Shaw}, S.~{Bharadwaj} and R.~{Mondal}, \emph{{The impact of
  non-Gaussianity on the error covariance for observations of the Epoch of
  Reionization 21-cm power spectrum}},
  \href{https://doi.org/10.1093/mnras/stz1561}{\emph{\mnras} {\bfseries 487}
  (2019) 4951} [\href{https://arxiv.org/abs/1902.08706}{{\ttfamily
  1902.08706}}].

\bibitem{shaw20}
A.K.~{Shaw}, S.~{Bharadwaj} and R.~{Mondal}, \emph{{The impact of
  non-Gaussianity on the Epoch of Reionization parameter forecast using 21-cm
  power-spectrum measurements}},
  \href{https://doi.org/10.1093/mnras/staa2090}{\emph{\mnras} {\bfseries 498}
  (2020) 1480} [\href{https://arxiv.org/abs/2005.06535}{{\ttfamily
  2005.06535}}].

\bibitem{majumdar18}
S.~{Majumdar}, J.R.~{Pritchard}, R.~{Mondal}, C.A.~{Watkinson}, S.~{Bharadwaj}
  and G.~{Mellema}, \emph{{Quantifying the non-Gaussianity in the EoR 21-cm
  signal through bispectrum}},
  \href{https://doi.org/10.1093/mnras/sty535}{\emph{\mnras} {\bfseries 476}
  (2018) 4007} [\href{https://arxiv.org/abs/1708.08458}{{\ttfamily
  1708.08458}}].

\bibitem{majumdar20}
S.~{Majumdar}, M.~{Kamran}, J.R.~{Pritchard}, R.~{Mondal}, A.~{Mazumdar},
  S.~{Bharadwaj} et~al., \emph{{Redshifted 21-cm bispectrum - I. Impact of the
  redshift space distortions on the signal from the Epoch of Reionization}},
  \href{https://doi.org/10.1093/mnras/staa3168}{\emph{\mnras} {\bfseries 499}
  (2020) 5090} [\href{https://arxiv.org/abs/2007.06584}{{\ttfamily
  2007.06584}}].

\bibitem{harker09}
G.J.A.~{Harker}, S.~{Zaroubi}, R.M.~{Thomas}, V.~{Jeli{\'c}}, P.~{Labropoulos},
  G.~{Mellema} et~al., \emph{{Detection and extraction of signals from the
  epoch of reionization using higher-order one-point statistics}},
  \href{https://doi.org/10.1111/j.1365-2966.2008.14209.x}{\emph{\mnras}
  {\bfseries 393} (2009) 1449}
  [\href{https://arxiv.org/abs/0809.2428}{{\ttfamily 0809.2428}}].

\bibitem{watkinson14}
C.A.~{Watkinson} and J.R.~{Pritchard}, \emph{{Distinguishing models of
  reionization using future radio observations of 21-cm 1-point statistics}},
  \href{https://doi.org/10.1093/mnras/stu1384}{\emph{\mnras} {\bfseries 443}
  (2014) 3090} [\href{https://arxiv.org/abs/1312.1342}{{\ttfamily 1312.1342}}].

\bibitem{shimabukuro15a}
H.~{Shimabukuro}, S.~{Yoshiura}, K.~{Takahashi}, S.~{Yokoyama} and K.~{Ichiki},
  \emph{{Studying 21cm power spectrum with one-point statistics}},
  \href{https://doi.org/10.1093/mnras/stv965}{\emph{\mnras} {\bfseries 451}
  (2015) 467} [\href{https://arxiv.org/abs/1412.3332}{{\ttfamily 1412.3332}}].

\bibitem{kubota16}
K.~{Kubota}, S.~{Yoshiura}, H.~{Shimabukuro} and K.~{Takahashi},
  \emph{{Expected constraints on models of the epoch of reionization with the
  variance and skewness in redshifted 21 cm-line fluctuations}},
  \href{https://doi.org/10.1093/pasj/psw059}{\emph{\pasj} {\bfseries 68} (2016)
  61} [\href{https://arxiv.org/abs/1602.02873}{{\ttfamily 1602.02873}}].

\bibitem{2019MNRAS.487.1101R}
H.E.~{Ross}, K.L.~{Dixon}, R.~{Ghara}, I.T.~{Iliev} and G.~{Mellema},
  \emph{{Evaluating the QSO contribution to the 21-cm signal from the Cosmic
  Dawn}}, \href{https://doi.org/10.1093/mnras/stz1220}{\emph{\mnras} {\bfseries
  487} (2019) 1101} [\href{https://arxiv.org/abs/1808.03287}{{\ttfamily
  1808.03287}}].

\bibitem{ross21}
H.E.~{Ross}, S.K.~{Giri}, G.~{Mellema}, K.L.~{Dixon}, R.~{Ghara} and
  I.T.~{Iliev}, \emph{{Redshift-space distortions in simulations of the 21-cm
  signal from the cosmic dawn}},
  \href{https://doi.org/10.1093/mnras/stab1822}{\emph{\mnras} {\bfseries 506}
  (2021) 3717} [\href{https://arxiv.org/abs/2011.03558}{{\ttfamily
  2011.03558}}].

\bibitem{shimabukuro15}
H.~{Shimabukuro}, S.~{Yoshiura}, K.~{Takahashi}, S.~{Yokoyama} and K.~{Ichiki},
  \emph{{21 cm line bispectrum as a method to probe cosmic dawn and epoch of
  reionization}}, \href{https://doi.org/10.1093/mnras/stw482}{\emph{\mnras}
  {\bfseries 458} (2016) 3003}
  [\href{https://arxiv.org/abs/1507.01335}{{\ttfamily 1507.01335}}].

\bibitem{watkinson17}
C.A.~{Watkinson}, S.~{Majumdar}, J.R.~{Pritchard} and R.~{Mondal}, \emph{{A
  fast estimator for the bispectrum and beyond - a practical method for
  measuring non-Gaussianity in 21-cm maps}},
  \href{https://doi.org/10.1093/mnras/stx2130}{\emph{\mnras} {\bfseries 472}
  (2017) 2436} [\href{https://arxiv.org/abs/1705.06284}{{\ttfamily
  1705.06284}}].

\bibitem{watkinson19}
C.A.~{Watkinson}, S.K.~{Giri}, H.E.~{Ross}, K.L.~{Dixon}, I.T.~{Iliev},
  G.~{Mellema} et~al., \emph{{The 21-cm bispectrum as a probe of
  non-Gaussianities due to X-ray heating}},
  \href{https://doi.org/10.1093/mnras/sty2740}{\emph{\mnras} {\bfseries 482}
  (2019) 2653} [\href{https://arxiv.org/abs/1808.02372}{{\ttfamily
  1808.02372}}].

\bibitem{ma21}
Q.-B.~{Ma}, B.~{Ciardi}, M.B.~{Eide}, P.~{Busch}, Y.~{Mao} and Q.-J.~{Zhi},
  \emph{{Investigating X-Ray Sources during the Epoch of Reionization with the
  21 cm Signal}}, \href{https://doi.org/10.3847/1538-4357/abefd5}{\emph{\apj}
  {\bfseries 912} (2021) 143}
  [\href{https://arxiv.org/abs/2103.09394}{{\ttfamily 2103.09394}}].

\bibitem{shimabukuro16b}
H.~{Shimabukuro}, S.~{Yoshiura}, K.~{Takahashi}, S.~{Yokoyama} and K.~{Ichiki},
  \emph{{Constraining the epoch-of-reionization model parameters with the 21-cm
  bispectrum}}, \href{https://doi.org/10.1093/mnras/stx530}{\emph{\mnras}
  {\bfseries 468} (2017) 1542}
  [\href{https://arxiv.org/abs/1608.00372}{{\ttfamily 1608.00372}}].

\bibitem{tiwari22}
H.~{Tiwari}, A.K.~{Shaw}, S.~{Majumdar}, M.~{Kamran} and M.~{Choudhury},
  \emph{{Improving constraints on the reionization parameters using 21-cm
  bispectrum}},
  \href{https://doi.org/10.1088/1475-7516/2022/04/045}{\emph{\jcap} {\bfseries
  2022} (2022) 045} [\href{https://arxiv.org/abs/2108.07279}{{\ttfamily
  2108.07279}}].

\bibitem{watkinson22}
C.A.~{Watkinson}, B.~{Greig} and A.~{Mesinger}, \emph{{Epoch of reionization
  parameter estimation with the 21-cm bispectrum}},
  \href{https://doi.org/10.1093/mnras/stab3706}{\emph{\mnras} {\bfseries 510}
  (2022) 3838} [\href{https://arxiv.org/abs/2102.02310}{{\ttfamily
  2102.02310}}].

\bibitem{hinshaw13}
G.~{Hinshaw}, D.~{Larson}, E.~{Komatsu}, D.N.~{Spergel}, C.L.~{Bennett},
  J.~{Dunkley} et~al., \emph{{Nine-year Wilkinson Microwave Anisotropy Probe
  (WMAP) Observations: Cosmological Parameter Results}},
  \href{https://doi.org/10.1088/0067-0049/208/2/19}{\emph{\apjs} {\bfseries
  208} (2013) 19} [\href{https://arxiv.org/abs/1212.5226}{{\ttfamily
  1212.5226}}].

\bibitem{planck14}
{Planck Collaboration}, P.A.R.~{Ade}, N.~{Aghanim}, C.~{Armitage-Caplan},
  M.~{Arnaud}, M.~{Ashdown} et~al., \emph{{Planck 2013 results. XVI.
  Cosmological parameters}},
  \href{https://doi.org/10.1051/0004-6361/201321591}{\emph{\aap} {\bfseries
  571} (2014) A16} [\href{https://arxiv.org/abs/1303.5076}{{\ttfamily
  1303.5076}}].

\bibitem{pritchard08}
J.R.~{Pritchard} and A.~{Loeb}, \emph{{Evolution of the 21cm signal throughout
  cosmic history}},
  \href{https://doi.org/10.1103/PhysRevD.78.103511}{\emph{\prd} {\bfseries 78}
  (2008) 103511} [\href{https://arxiv.org/abs/0802.2102}{{\ttfamily
  0802.2102}}].

\bibitem{field58}
G.B.~{Field}, \emph{{Excitation of the Hydrogen 21-CM Line}},
  \href{https://doi.org/10.1109/JRPROC.1958.286741}{\emph{Proceedings of the
  IRE} {\bfseries 46} (1958) 240}.

\bibitem{2002ApJ...572L.123I}
I.T.~{Iliev}, P.R.~{Shapiro}, A.~{Ferrara} and H.~{Martel}, \emph{{On the
  Direct Detectability of the Cosmic Dark Ages: 21 Centimeter Emission from
  Minihalos}}, \href{https://doi.org/10.1086/341869}{\emph{\apjl} {\bfseries
  572} (2002) L123} [\href{https://arxiv.org/abs/astro-ph/0202410}{{\ttfamily
  astro-ph/0202410}}].

\bibitem{2020MNRAS.498.4178M}
R.~{Mondal}, A.~{Fialkov}, C.~{Fling}, I.T.~{Iliev}, R.~{Barkana}, B.~{Ciardi}
  et~al., \emph{{Tight constraints on the excess radio background at z = 9.1
  from LOFAR}}, \href{https://doi.org/10.1093/mnras/staa2422}{\emph{\mnras}
  {\bfseries 498} (2020) 4178}
  [\href{https://arxiv.org/abs/2004.00678}{{\ttfamily 2004.00678}}].

\bibitem{2021MNRAS.503.4551G}
R.~{Ghara}, S.K.~{Giri}, B.~{Ciardi}, G.~{Mellema} and S.~{Zaroubi},
  \emph{{Constraining the state of the intergalactic medium during the Epoch of
  Reionization using MWA 21-cm signal observations}},
  \href{https://doi.org/10.1093/mnras/stab776}{\emph{\mnras} {\bfseries 503}
  (2021) 4551} [\href{https://arxiv.org/abs/2103.07483}{{\ttfamily
  2103.07483}}].

\bibitem{2022JCAP...03..055G}
R.~{Ghara}, G.~{Mellema} and S.~{Zaroubi}, \emph{{Astrophysical information
  from the Rayleigh-Jeans Tail of the CMB}},
  \href{https://doi.org/10.1088/1475-7516/2022/03/055}{\emph{\jcap} {\bfseries
  2022} (2022) 055} [\href{https://arxiv.org/abs/2108.13593}{{\ttfamily
  2108.13593}}].

\bibitem{ghara15}
R.~{Ghara}, K.K.~{Datta} and T.R.~{Choudhury}, \emph{{21 cm signal from cosmic
  dawn - II. Imprints of the light-cone effects}},
  \href{https://doi.org/10.1093/mnras/stv1855}{\emph{\mnras} {\bfseries 453}
  (2015) 3143} [\href{https://arxiv.org/abs/1504.05601}{{\ttfamily
  1504.05601}}].

\bibitem{ghara18}
R.~{Ghara}, G.~{Mellema}, S.K.~{Giri}, T.R.~{Choudhury}, K.K.~{Datta} and
  S.~{Majumdar}, \emph{{Prediction of the 21-cm signal from reionization:
  comparison between 3D and 1D radiative transfer schemes}},
  \href{https://doi.org/10.1093/mnras/sty314}{\emph{\mnras} {\bfseries 476}
  (2018) 1741} [\href{https://arxiv.org/abs/1710.09397}{{\ttfamily
  1710.09397}}].

\bibitem{2019JCAP...02..058G}
S.K.~{Giri}, A.~{D'Aloisio}, G.~{Mellema}, E.~{Komatsu}, R.~{Ghara} and
  S.~{Majumdar}, \emph{{Position-dependent power spectra of the 21-cm signal
  from the epoch of reionization}},
  \href{https://doi.org/10.1088/1475-7516/2019/02/058}{\emph{\jcap} {\bfseries
  2019} (2019) 058} [\href{https://arxiv.org/abs/1811.09633}{{\ttfamily
  1811.09633}}].

\bibitem{Harnois12}
J.~{Harnois-D{\'e}raps}, U.-L.~{Pen}, I.T.~{Iliev}, H.~{Merz}, J.D.~{Emberson}
  and V.~{Desjacques}, \emph{{High-performance P$^{3}$M N-body code:
  CUBEP$^{3}$M}}, \href{https://doi.org/10.1093/mnras/stt1591}{\emph{\mnras}
  {\bfseries 436} (2013) 540}.

\bibitem{behroozi15}
P.S.~Behroozi and J.~Silk, \emph{A {SIMPLE} {TECHNIQUE} {FOR} {PREDICTING}
  {HIGH}-{REDSHIFT} {GALAXY} {EVOLUTION}},
  \href{https://doi.org/10.1088/0004-637x/799/1/32}{\emph{The Astrophysical
  Journal} {\bfseries 799} (2015) 32}.

\bibitem{sun16}
G.~Sun and S.R.~Furlanetto, \emph{Constraints on the star formation efficiency
  of galaxies during the epoch of reionization}, {\emph{Monthly Notices of the
  Royal Astronomical Society} {\bfseries 460} (2016) 417}.

\bibitem{Fioc97}
M.~{Fioc} and B.~{Rocca-Volmerange}, \emph{{PEGASE: a UV to NIR spectral
  evolution model of galaxies. Application to the calibration of bright galaxy
  counts.}}, {\emph{\aap} {\bfseries 326} (1997) 950}
  [\href{https://arxiv.org/abs/astro-ph/9707017}{{\ttfamily
  astro-ph/9707017}}].

\bibitem{2019MNRAS.487.2785I}
N.~{Islam}, R.~{Ghara}, B.~{Paul}, T.R.~{Choudhury} and B.B.~{Nath},
  \emph{{Cosmological implications of the composite spectra of galactic X-ray
  binaries constructed using MAXI data}},
  \href{https://doi.org/10.1093/mnras/stz1446}{\emph{\mnras} {\bfseries 487}
  (2019) 2785} [\href{https://arxiv.org/abs/1905.10386}{{\ttfamily
  1905.10386}}].

\bibitem{ghara20}
R.~{Ghara}, S.K.~{Giri}, G.~{Mellema}, B.~{Ciardi}, S.~{Zaroubi}, I.T.~{Iliev}
  et~al., \emph{{Constraining the intergalactic medium at z
  {\ensuremath{\approx}} 9.1 using LOFAR Epoch of Reionization observations}},
  \href{https://doi.org/10.1093/mnras/staa487}{\emph{\mnras} (2020) }
  [\href{https://arxiv.org/abs/2002.07195}{{\ttfamily 2002.07195}}].

\bibitem{barkana04}
R.~Barkana and A.~Loeb, \emph{Unusually large fluctuations in the statistics of
  galaxy formation at high redshift}, {\emph{The Astrophysical Journal}
  {\bfseries 609} (2004) 474}.

\bibitem{ghara15a}
R.~{Ghara}, T.R.~{Choudhury} and K.K.~{Datta}, \emph{{21 cm signal from cosmic
  dawn: imprints of spin temperature fluctuations and peculiar velocities}},
  \href{https://doi.org/10.1093/mnras/stu2512}{\emph{\mnras} {\bfseries 447}
  (2015) 1806} [\href{https://arxiv.org/abs/1406.4157}{{\ttfamily 1406.4157}}].

\bibitem{2020MNRAS.494.3294N}
A.~{Nasirudin}, I.T.~{Iliev} and K.~{Ahn}, \emph{{Modelling the stochasticity
  of high-redshift halo bias}},
  \href{https://doi.org/10.1093/mnras/staa853}{\emph{\mnras} {\bfseries 494}
  (2020) 3294} [\href{https://arxiv.org/abs/1910.12452}{{\ttfamily
  1910.12452}}].

\bibitem{bharadwaj20}
S.~Bharadwaj, A.~Mazumdar and D.~Sarkar, \emph{{Quantifying the redshift space
  distortion of the bispectrum I: primordial non-Gaussianity}},
  \href{https://doi.org/10.1093/mnras/staa279}{\emph{\mnras} {\bfseries 493}
  (2020) 594}
  [\href{https://arxiv.org/abs/https://academic.oup.com/mnras/article-pdf/493/1/594/32513251/staa279.pdf}{{\ttfamily
  https://academic.oup.com/mnras/article-pdf/493/1/594/32513251/staa279.pdf}}].

\bibitem{watkinson20}
C.A.~Watkinson, C.M.~Trott and I.~Hothi, \emph{{The bispectrum and 21-cm
  foregrounds during the Epoch of Reionization}},
  \href{https://doi.org/10.1093/mnras/staa3677}{\emph{Monthly Notices of the
  Royal Astronomical Society} {\bfseries 501} (2020) 367}
  [\href{https://arxiv.org/abs/https://academic.oup.com/mnras/article-pdf/501/1/367/34949789/staa3677.pdf}{{\ttfamily
  https://academic.oup.com/mnras/article-pdf/501/1/367/34949789/staa3677.pdf}}].

\bibitem{hothi20}
I.~Hothi, E.~Chapman, J.R.~Pritchard, F.G.~Mertens, L.V.E.~Koopmans, B.~Ciardi
  et~al., \emph{{Comparing foreground removal techniques for recovery of the
  LOFAR-EoR 21 cm power spectrum}},
  \href{https://doi.org/10.1093/mnras/staa3446}{\emph{Monthly Notices of the
  Royal Astronomical Society} {\bfseries 500} (2020) 2264}
  [\href{https://arxiv.org/abs/https://academic.oup.com/mnras/article-pdf/500/2/2264/34497733/staa3446.pdf}{{\ttfamily
  https://academic.oup.com/mnras/article-pdf/500/2/2264/34497733/staa3446.pdf}}].

\bibitem{yoshiura15}
S.~{Yoshiura}, H.~{Shimabukuro}, K.~{Takahashi}, R.~{Momose}, H.~{Nakanishi}
  and H.~{Imai}, \emph{{Sensitivity for 21 cm bispectrum from Epoch of
  Reionization}}, \href{https://doi.org/10.1093/mnras/stv855}{\emph{\mnras}
  {\bfseries 451} (2015) 266}
  [\href{https://arxiv.org/abs/1412.5279}{{\ttfamily 1412.5279}}].

\end{thebibliography}\endgroup



\appendix

\end{document}